\documentstyle[aps,multicol]{revtex}
\begin{document}
\title{The sudden approximation in photoemission and beyond}
\author{L. Hedin (1,2) and J. D. Lee (3)}
\address{1. Dept. of Physics, University of Lund, S\"{o}lvegatan 14A, 22362 Lund,\\
Sweden\\
2. MPI-FKF, Heisenbergstrasse 1, D-70569 Stuttgart, Germany\\
3. Dept. of Physics and Dept. of Complexity Science and\\
Engineering,\thinspace University of Tokyo, Bunkyo ku, Tokyo 113,\\
Japan}
\date{\today}
\maketitle

\begin{abstract}
The conditions behind the sudden approximation are critically examined. The
fuzzy band expression is derived in detail from first principles. We go
beyond the sudden approximation to account for both extrinsic losses and
interference effects. In an extension of earlier work we discuss both core
and valence satellites including both intrinsic and extrinsic amplitudes,
and high energy excitations as well as low energy electron-hole pairs. We
show how the extrinsic losses in photoemission can be connected with the
electron energy loss function. This is achieved by three approximations, to
connect the dynamically screened potential in the bulk solid to the loss
function, to account for the surface, and lastly to extrapolate zero
momentum loss data to include dispersion. The extrinsic losses are found to
be weak for small loss energies. For larger loss energies the extrinsic
losses can be strong and have strong interference with the intrinsic losses
depending on the nature of the solid. The transition from the adiabatic to
the sudden regime is discussed for atoms and localized systems and compared
with the situation for solids. We argue that in solids for photoelectron
energies above some 10 eV the external losses are mainly connected with
spacially extended excitations. We discuss strongly correlated
quasi-two-dimensional solids with Bi2212 as example, and find an asymmetric
broadening of the main peak from shake up of acoustic three-dimensional
plasmons. In the superconducting state the loss function is assumed to have
a gap, which then leads to a peak-dip-hump structure in the photoemission
spectrum. This structure is compared with a corresponding structure in
tunneling.
\end{abstract}

\begin{multicols}{2}

\narrowtext

\section{Introduction }

Photoemission spectroscopy (PES) has become an important tool to analyze
electronic structure. The accuracy of the experimental energy resolution is
now in the meV range, and sample control has reached a high level of
reliability. Yet the theory used to interpret these high quality
experimental results has great weaknesses. Often only the spectral function
is compared to experiment. This means that e.g. the momentum dependence of
the dipole matrix elements is neglected, as well as the extrinsic losses.
Use of the spectral function is fine if we have well-defined sharp
quasi-particle peaks, and want to do band structure mapping. As the
experimental resolution increases and line shapes are resolved this level of
approximation becomes increasingly doubtful. That is in particular true for
strongly correlated systems like high temperature superconductors (HTC)
which have important features that cannot be described by quasi-particle
peaks.

Presently PES for HTC materials is often compared to spectral functions for
models like the t-J one for a two-dimensional (2D) copper oxide layer. Such
a model can only (at best) describe a narrow low energy region of the total
excitation spectrum. The omitted degrees of freedom describe higher
energies, and also collective low energy modes with three-dimensional
character. As far as we know, there has been no systematic attempt to
develop a theory that both gives a good description of the strongly
correlated system and also takes account of the remaining correlations. This
is an important and urgent problem if we want to have a true a priori
understanding of e.g. HTC materials. In lack of a systematic theory we
sketch below an approach to this problem built on physical intuition.

The 2D problem is normally treated with a model Hamiltonian based on a small
number of orbitals $\{\phi _{2D}\}$. The parameters in the model Hamiltonian
can be obtained from ''down-folding'' the basis functions in a bandstructure
calculation by using e.g. developments of the very efficient LMTO formalism 
\cite{Andersen00}. The interesting correlation effects come from the
degeneracy in the problem, i.e. the large number of possibilities to occupy
the different $m_{l}$ and $m_{s}$ states of the localized muffin tin
orbitals connected with the down-folded part of the bandstructure. There are
also efficient methods based on the LMTO formalism to handle a wide energy
range which treats valence and conduction bands on an equal footing so that
the wave functions are orthogonal \cite{Aryasetiawan94}. Let's call the
additional set of states beyond $\{\phi _{2D}\}$ for $\{\phi _{3D}\}$. We
assume that $\{\phi _{3D}\}$ is strictly orthogonal to $\{\phi _{2D}\}$
despite any complications that can come from down-folding. The one electron
states $\{\phi _{2D}\}$ give rise to a huge number of configurations which
then all are orthogonal to the configurations that can be built from the set 
$\{\phi _{3D}\}$. We also assume that we can write the state vectors as a
direct product, $\left| \Psi _{tot},s_{1},s_{2}\right\rangle =\left| \Psi
_{2D},s_{1}\right\rangle \left| \Psi _{3D}^{(s_{1})},s_{2}\right\rangle $,
where both the 2D and 3D parts are correlated state vectors. A parametric
dependence of the 3D state on the 2D state is obvious when we have a core
hole in the 2D system, but is actually important also in general.

When we develop our theory we will not make any specific assumptions about
the correlations in the 2D system. We will assume that the excitations in
the 3D system are extended in space and can be described by a quasi-boson
model. This is possible if we only want to describe PES for energies larger
than about 10 eV where a strongly correlated localized system already has
reached the sudden limit. \cite{Lee99} The treatment here of strongly
correlated systems is based on our recent work. \cite{HL01}

\section{The ''one-step'' theory}

Bandstructure and surface effects in photoemission are today often treated
in the so called ''one-step model'' pioneered by Pendry \cite{Pendry76}, 
\cite{Pendry78}, for recent reviews see e. g. Refs. \cite{Inglesfield92}, 
\cite{Schattke97}, \cite{Gunnarsson01}, \cite{Kevan01}. In the one-step
model the important dipole matrix elements are included, as well as surface
photoemission (emission due to the gradient in the electromagnetic field,
for recent work see e. g. Ref. \cite{Claesson99}), but not the extrinsic
losses.

A complete theory was developed by Caroli et al in 1973 \cite{Caroli73}
based on the three current expression of Schaich and Ashcroft. \cite
{Schaich71} The Caroli et al theory uses the Keldysh formalism \cite
{Keldysh65} for non-equilibrium Green's functions. It is exact and one can e
g identify the diagrams which give extrinsic losses. Caroli et al did not
however extract any practical calculation scheme from their theory.

Important steps towards a practical photoemission theory were taken by
Feibelman and Eastman \cite{Feibelman74} and by Pendry \cite{Pendry76}, \cite
{Pendry78}, who discussed the lowest non-trivial diagram in the Caroli et al
expansion, assuming the final states to be (time-inverted) LEED scattering
states. Pendry and collaborators developed an explicit multiple scattering
formulation and computer programs rooted in earlier work on the LEED
problem. \cite{Pendry80} A careful discussion of the Caroli et al theory and
the basics of photoemission was given by Almbladh 1985 leading to some
explicit approximations. \cite{Almbladh85}

With todays precise experimental data on photoemission the need for a
comprehensive and practical theory is great. The one-step theory has serious
limitations both in its theoretical foundation and in its scope. Thus
Pendry, Feibelman and Eastman just assumed that the photoelectron states
should be damped. That this is correct was shown later by Almbladh, however
he uses the Keldysh diagram expansion which is rather involved. More serious
is that the one-step theory neglects the extrinsic losses. These losses were
early recognized to be important, and in core electron photoemission they
are routinely included. That they also are important in photoemission from
an extended state (valence electrons) is not as widely recognized.

A full theory for photoemission has a number of intriguing features. We will
display some of them by starting from the Golden Rule expression rather than
the Keldysh expansion, and postulating a simple model Hamiltonian. Exact
results can then be obtained by elementary means. The model Hamiltonian will
be motivated a posteriori by discussing cases where the results can be
obtained by other means.

The Golden Rule expression for the photo-electron current can be obtained
from scattering theory \cite{GW64} (we use atomic units, $e=\hbar =m=1$, and
thus e.g. energies are in Hartrees, 27.2 eV),

\begin{equation}
J_{{\bf k}}\left( \omega \right) =\sum_{s}\left| \left\langle N-1,s;{\bf k}%
\left| \Delta \right| N\right\rangle \right| ^{2}\delta \left( \omega
-\varepsilon _{{\bf k}}+\varepsilon _{s}\right) .  \label{Jk}
\end{equation}
The final states $\left| N-1,s;{\bf k}\right\rangle $ have the proper
boundary conditions for scattering states. All states are correlated.
Further $\varepsilon _{{\bf k}}$ is the energy of the photo-electron, $%
\varepsilon _{{\bf k}}={\bf k}^{2}/2\;$\cite{note1}, $\varepsilon _{s}$
gives the energy of the final state of the solid as $\varepsilon
_{s}=E\left( N,0\right) -E\left( N-1,s\right) $, $\omega $ is the photon
energy, and $\Delta $ the optical transition operator, 
\[
\Delta =\sum_{ij}\Delta _{ij}c_{i}^{\dagger }c_{j}^{{}},\;\Delta
_{ij}=\left\langle i\left| Ap+pA\right| j\right\rangle . 
\]
$\left| N\right\rangle $ is the initial state. The exact expression for $J_{%
{\bf k}}\left( \omega \right) $ is rather intractable, and we have to
develop approximations that can be handled. \cite{Bardy85}, \cite{Hedin98}, 
\cite{Hedin99} For non-interacting electrons where $\left| N-1,s;{\bf k}%
\right\rangle $ and $\left| N\right\rangle $ $\ $are Slater determinants, it
is easy to see that $\left| N-1,s;{\bf k}\right\rangle =c_{{\bf k}}^{\dagger
}c_{s}^{{}}\left| N\right) $, and thus Eq. \ref{Jk} reduces to the
well-known form 
\[
J_{{\bf k}}\left( \omega \right) =\sum_{s}^{occ}\left| \Delta _{{\bf k}%
s}\right| ^{2}\delta \left( \omega -\varepsilon _{{\bf k}}+\varepsilon
_{s}\right) , 
\]
where the index $"s"$ now stands for an occupied one-electron state (e.g. a
core electron state or a state in the valence band).

The final state can be written as,

\begin{equation}
\left| N-1,s;{\bf k}\right\rangle =\left[ 1+\frac{1}{E-H-i\eta }\left(
H-E\right) \right] c_{{\bf k}}^{\dagger }\left| N-1,s\right\rangle ,
\label{Psif}
\end{equation}
where $H$ is the fully interacting Hamiltonian including target electrons
and photoelectron, and $E=\omega +E\left( N,0\right) =\varepsilon _{{\bf k}%
}+E\left( N-1,s\right) .$ Further $\left| N-1,s\right\rangle $ is the target
state after the photo-electron has left, and $c_{{\bf k}}^{\dagger }$
creates the photoelectron. The states corresponding to $c_{{\bf k}}^{\dagger
}$ are so far undefined, except that they are time-inverted LEED states with
asymptotically a plane wave part $\exp (i{\bf k}\cdot {\bf r})$. Usually in
scattering problems $\left( H-E\right) $ is replaced by the interaction $V$
between the scattering particle and the target. This is not possible when
the scattering particle is identical with particles in the target. One way
to avoid this difficulty is to regard the scattering particle as different,
and do an anti-symmetrization of the solution at the end of the analysis,
which is the route followed in the standard treatise of scattering by
Goldberger and Watson. \cite{GW64}

At high energies we can regard the scattering particle as different from the
particles in the target since the exchange coupling becomes negligible. We
can then replace $\left( H-E\right) $ by a potential $V$ which describes the
coupling between the photoelectron and the density fluctuations in the
solid. \cite{Hedin98} This is true for all systems, also strongly correlated
ones. For strongly correlated {\it localized} systems the interaction with
the photoelectron becomes small already at energies of about 10 eV. \cite
{Lee99} \ As discussed in the introduction we will take the total wave
function as a direct product, $\left| \Psi _{tot},s_{1},s_{2}\right\rangle
=\left| \Psi _{2D},s_{1}\right\rangle \left| \Psi
_{3D}^{(s_{1})},s_{2}\right\rangle $. For energies above 10 eV we only need
to consider the interaction between the photoelectron and the excitations $%
"s_{2}"$.

If the $(H-E)$ term, which gives the extrinsic scattering, is neglected, and
if we assume that $c_{{\bf k}}^{{}}\left| N\right\rangle =0$, the amplitude
becomes 
\begin{equation}
\left\langle N-1,s;{\bf k}\left| \Delta \right| N\right\rangle
=\sum_{j}\left\langle {\bf k}\left| \Delta \right| j\right\rangle
\left\langle N-1,s\left| c_{j}\right| N\right\rangle .  \label{tau0}
\end{equation}
The photo current in Eq. \ref{Jk} is then given by the spectral function $%
A_{ij}\left( \varepsilon _{{\bf k}}-\omega \right) $, 
\begin{equation}
J_{{\bf k}}(\omega )=\sum_{ij}\left\langle {\bf k}\left| \Delta \right|
i\right\rangle A_{ij}\left( \varepsilon _{{\bf k}}-\omega \right)
\left\langle j\left| \Delta \right| {\bf k}\right\rangle ,  \label{Jk2}
\end{equation}
where 
\begin{eqnarray}\label{Aij}
A_{ij}\left( \omega \right) &=&\sum_{s}^{\varepsilon _{s}<\mu }\left\langle
N\left| c_{i}^{\dagger }\right| N-1,s\right\rangle \left\langle N-1,s\left|
c_{j}^{{}}\right| N\right\rangle \nonumber \\
&& \ \ \ \ \ \times\delta \left( \omega -\varepsilon
_{s}\right) . 
\end{eqnarray}
Eq. \ref{Jk2} is the well-known intrinsic approximation for the
photocurrent. The photoelectron states $\left| {\bf k}\right\rangle $ are
time-inverted LEED\ states (with no magnetic field this just means complex
conjugation). A LEED state has an incoming and a set of scattered plane
waves outside the solid, matched to a solution inside the solid with a
potential $V_{H}+\Sigma \left( \varepsilon _{{\bf k}}\right) $, where $%
V_{H}\left( {\bf r}\right) $ is the Hartree potential for a finite solid
with a surface and $\Sigma $ the complex selfenergy. In the next section we
will show how the perturbation $V$ gives rise to extrinsic energy losses and
the selfenergy term $\Sigma \left( \varepsilon _{{\bf k}}\right) $.

In the one-electron approximation the spectral function is diagonal and has
just one quasi-particle peak, $A_{ij}\left( \omega \right) =\delta \left(
\omega -E_{i}\right) \delta _{ij}$. In most applications the spectral
function is taken as diagonal. This is e g generally justified in the GW
approximation. \cite{Hedin99} A core electron spectral function for a metal $%
A_{c}\left( \omega \right) $ has an asymmetric peak (MND singularity \cite
{Mahan81}) followed by a set of plasmon peaks. The asymmetry is due to
shake-up of electron-hole pairs and phonons. \cite{Citrin77} For a
conduction electron in a metal the spectral function $A_{{\bf k}}\left(
\omega \right) $ looks quite similar with a broadened and asymmetric
quasi-particle peak and a pronounced satellite structure. The broadening and
asymmetry however go to zero at the Fermi level. Semiconductors have no low
energy electron-hole excitations and thus lack the MND singularity, but
still have pronounced plasmon satellites. Transition metals have about the
same total strength in the satellite structure, but lack the pronounced high
energy peaks of sp-metals and valence semiconductors. Strongly correlated
materials may lack the quasi-particle peak altogether, and have all the
spectral strength in the satellite.

For core electron photoemission the current in the intrinsic approximation
is simply $J_{{\bf k}}(\omega )=\left| \left\langle {\bf k}\left| \Delta
\right| c\right\rangle \right| ^{2}A_{c}\left( \varepsilon _{{\bf k}}-\omega
\right) $. For valence electrons we need a more involved discussion. We
denote the initial state $"j"$ in Eq. \ref{tau0} with a vector ${\bf l}$
which also includes a bandindex, and Eq. \ref{Jk2} becomes 
\begin{equation}
J_{{\bf k}}(\omega )=\sum_{{\bf l}}\left| \left\langle {\bf k}\left| \Delta
\right| {\bf l}\right\rangle \right| ^{2}A_{{\bf l}}\left( \varepsilon _{%
{\bf k}}-\omega \right) .  \label{Jk4}
\end{equation}
For energies close to the quasi-particle energy $E_{{\bf l}}$, the spectral
function $A_{{\bf l}}$ has the form (neglecting the asymmetry, which can be
pronounced away from the Fermi level) 
\begin{equation}
A_{{\bf l}}\left( \omega \right) =\frac{Z_{{\bf l}}}{\pi }\frac{\Gamma _{%
{\bf l}}}{\left( \omega -E_{{\bf l}}\right) ^{2}+\Gamma _{{\bf l}}^{2}},
\label{Al}
\end{equation}
which together with Eq. \ref{Jk4} gives the one-step approximation. The
photoelectron wavefunction $\left| {\bf k}\right\rangle $ inside the solid
is determined by the potential $V_{H}\left( {\bf r}\right) +\Sigma \left( 
{\bf r},{\bf r}^{\prime };\varepsilon _{{\bf k}}\right) $, where $V_{H}$ is
the Hartree or Coulomb potential. The real part of the non-local optical
potential or selfenergy $\Sigma $ is usually approximated by a local
potential and combined with $V_{H}$ to give an effective potential $V_{eff}$%
. Replacing $-%
\mathop{\rm Im}%
\Sigma $ by its expectation value $\Gamma _{f}>0$, and setting $\left| {\bf k%
}\right\rangle =\left| \widetilde{{\bf k}}\right\rangle $ inside the solid,
we have 
\begin{equation}
\left( t+V_{eff}-i\Gamma _{f}\right) \left| \widetilde{{\bf k}}\right\rangle
=\varepsilon _{{\bf k}}\left| \widetilde{{\bf k}}\right\rangle ,
\label{heff}
\end{equation}
where $t$ is the kinetic energy, $\Gamma _{f}=-\left\langle \widetilde{{\bf k%
}}\left| 
\mathop{\rm Im}%
\Sigma \left( \varepsilon _{{\bf k}}\right) \right| \widetilde{{\bf k}}%
\right\rangle >0$ and $\varepsilon _{{\bf k}}={\bf k}^{2}/2$. \ The
photoelectron wavefunction $\left| \widetilde{{\bf k}}\right\rangle $ inside
the solid can be written as a combination of Bloch functions with complex $%
{\bf k}$ vectors, such that it correctly matches the plane waves outside the
solid. If one Bloch wave dominates \cite{Spanjaard77} its complex momentum $%
\widetilde{{\bf k}}$ follows from 
\[
E_{_{\widetilde{{\bf k}}}}=\varepsilon _{{\bf k}}+i\Gamma _{f},
\]
where $E_{_{\widetilde{{\bf k}}}}$\ is the complex quasiparticle energy
having both a shift in its real part from $%
\mathop{\rm Re}%
\Sigma $ and an imaginary part $\Gamma _{f}$. We write $\widetilde{{\bf k}}%
{\bf =}\widetilde{{\bf k}}_{\Vert }+\widehat{z}\left( \widetilde{k}%
_{z}+ik_{f}^{I}\right) $, where $\widetilde{{\bf k}}_{\Vert }$ is parallel
to the surface and $\widehat{z}$ is a unit vector normal to the surface. Due
to momentum conservation we have $\widetilde{{\bf k}}_{\Vert }={\bf k}%
_{\Vert }$. In the limit of small $\Gamma _{f}$ we can expand $E_{_{%
\widetilde{{\bf k}}}}=E_{%
\mathop{\rm Re}%
\widetilde{{\bf k}}}+$ $i%
\mathop{\rm Im}%
\widetilde{{\bf k}}\;\partial E_{_{\widetilde{{\bf k}}}}/\partial \widetilde{%
{\bf k}}$, and thus we have 
\[
k_{f}^{I}=\frac{\Gamma _{f}}{\partial E_{_{\widetilde{{\bf k}}}}/\partial 
\widetilde{k}_{z}}.
\]
Mahan \cite{Mahan70} and in more detail Spanjaard et al \cite{Spanjaard77}
studied the matrix element $\left\langle \widetilde{{\bf k}}\left| \Delta
\right| {\bf l}\right\rangle $ in the limit when the damping is small. The
integration parallel to the surface gives momentum conservation, $\widetilde{%
{\bf k}}_{\Vert }={\bf l}_{\Vert }$. The integration in the $z$ direction
gives a surface contribution plus contributions from the unit cells where
there is bulk symmetry. With bulk symmetry all cells give the same
contribution but for a factor, and we can sum to have 
\begin{eqnarray}\label{eq:matrixelement}
& &\left\langle \widetilde{{\bf k}}\left| \Delta \right| {\bf l}\right\rangle
\\ \nonumber
&=&\left\{ \left\langle \widetilde{{\bf k}}\left| \Delta \right| {\bf l}%
\right\rangle _{surf}+\frac{\left\langle \widetilde{{\bf k}}\left| \Delta
\right| {\bf l}\right\rangle _{0}}{1-\exp \left[ i\left( l_{z}-\widetilde{k}%
_{z}\right) c-k_{f}^{I}c\right] }\right\} \delta _{\widetilde{{\bf k}}%
_{\Vert },{\bf l}_{\Vert }},
\end{eqnarray}
where $c$ is the cell length in the $z$ direction and $\left\langle 
\widetilde{{\bf k}}\left| \Delta \right| {\bf l}\right\rangle _{0}$ is the
integral over the first (closest to the surface) bulk unit cell. Band
indices and 2D reciprocal lattice vectors have to be supplemented where
appropriate. With small $k_{f}^{I}$ and neglecting the surface contribution $%
\left| \left\langle \widetilde{{\bf k}}\left| \Delta \right| {\bf l}%
\right\rangle \right| ^{2}$ is sharply peaked as function of $l_{z}$ around $%
l_{z}^{0}$, 
\[
l_{z}^{0}=%
\mathop{\rm mod}%
(\widetilde{k}_{z},2\pi /c),
\]
and one can approximate 
\[
\left| \left\langle \widetilde{{\bf k}}\left| \Delta \right| {\bf l}%
\right\rangle \right| ^{2}=\frac{1}{c^{2}}\frac{\left| \left\langle 
\widetilde{{\bf k}}\left| \Delta \right| {\bf l}^{0}\right\rangle
_{0}\right| ^{2}}{\left( l_{z}^{0}-l_{z}\right) ^{2}+\left( k_{f}^{I}\right)
^{2}}\delta _{\widetilde{{\bf k}}_{\Vert },{\bf l}_{\Vert }},
\]
where $\left| {\bf l}^{0}\right\rangle $ is a wavefunction in the occupied
valence band, and 
\[
{\bf l}^{0}={\bf k}_{\Vert }+\widehat{z}l_{z}^{0}={\bf l}^{0}\left( {\bf k}%
\right) .
\]
The surface contribution to the photocurrent can be comparable to the bulk
one, and interference between the two terms in Eq. \ref{eq:matrixelement} is
then important. \cite{Claesson99}, \cite{Miller01} Collecting our results
the contribution to the photocurrent from the quasi-particle peak is 
\begin{eqnarray*}
&&J_{{\bf k}}\left( \omega \right)
\\ \nonumber
&=&\sum_{{\bf l}}\delta _{\widetilde{{\bf k}}%
_{\Vert },{\bf l}_{\Vert }}\frac{Z_{{\bf l}}}{\pi c^{2}}\frac{\Gamma _{i}}{%
\left( \varepsilon _{_{{\bf k}}}-E_{{\bf l}}-\omega \right) ^{2}+\Gamma
_{i}^{2}}\frac{\left| \left\langle \widetilde{{\bf k}}\left| \Delta \right| 
{\bf l}^{0}\right\rangle _{0}\right| ^{2}}{\left( l_{z}^{0}-l_{z}\right)
^{2}+\left( k_{f}^{I}\right) ^{2}},
\end{eqnarray*}
where $\Gamma _{i}=\Gamma _{{\bf l}}$.

With $\Gamma _{i}$ small we can expand $\varepsilon _{_{{\bf k}}}-E_{{\bf l}%
}-\omega =-\left( l_{z}-l_{z}^{1}\right) \partial E_{{\bf l}}/\partial l_{z}$%
, where $l_{z}^{1}$ is given by the resonance condition 
\begin{equation}
\varepsilon _{_{{\bf k}}}-E_{{\bf k}_{\Vert }+\widehat{z}l_{z}^{1}}-\omega
=0.  \label{deflz}
\end{equation}
We define 
\[
{\bf l}^{1}={\bf k}_{\Vert }+\widehat{z}l_{z}^{1}={\bf l}^{1}\left( {\bf k}%
,\omega \right) .
\]
When also $\Gamma _{i}$ is small we can integrate over $l_{z}$ keeping
everything constant except the resonance denominators. We then have a
convolution of two Lorentzians, which gives a new Lorentzian with the sum of
the widths 
\begin{eqnarray} \label{Jk5}
&&J_{{\bf k}}\left( \omega \right)
\\ \nonumber 
&\sim& \left( \frac{1}{k_{i}^{I}}+\frac{1}{%
k_{f}^{I}}\right) \left| \frac{\partial E_{{\bf l}^{1}}}{\partial l_{z}}%
\right| ^{-2}\frac{1}{c^{2}}\frac{Z_{{\bf l}^{1}}\Gamma _{i^{{}}}\left|
\left\langle \widetilde{{\bf k}}\left| \Delta \right| {\bf l}%
^{0}\right\rangle _{0}\right| ^{2}}{\left( l_{z}^{0}-l_{z}^{1}\right)
^{2}+\left( k_{i}^{I}+k_{f}^{I}\right) ^{2}}, 
\end{eqnarray}
where 
\[
k_{i}^{I}=\left( \frac{\Gamma _{i}}{\partial E_{{\bf l}}/\partial l_{z}}%
\right) _{{\bf l}={\bf l}^{0}},\;k_{f}^{I}=\left( \frac{\Gamma _{f}}{%
\partial E_{{\bf k}}/\partial k_{z}}\right) _{{\bf k=%
\mathop{\rm Re}%
}\widetilde{{\bf k}}}.
\]
The momentum parallel to the surface is conserved, $\widetilde{{\bf k}}%
_{\Vert }={\bf k}_{\Vert }$, and thus immediately follows from experiment
while the determination of $\widetilde{k}_{z}$ requires knowledge of $V_{eff}
$ and thus e.g. $%
\mathop{\rm Re}%
\Sigma $. There are however more involved experimental techniques to obtain
also $\widetilde{k}_{z}$. \cite{Hüfner99} As before band indices and 2D
reciprocal lattice vectors have to be supplemented where appropriate.

The transmission problem related to the matching of the photoelectron
wavefunction inside the solid and the plane waves outside has been studied
by Spanjaard et al. \cite{Spanjaard77} who conclude that for Ag the
transmission factor is about 0.5 at all angles and for all energies of
interest. The connection of the ''fuzzy band'' expression and different
modes of experiment has been discussed in detail by Smith et al. \cite
{Smith93} \ The deviation from free electron energies for the higher bands,
which shows up as the difference between $\widetilde{{\bf k}}$ and ${\bf k}$%
, persist quite high up in energy. \cite{Strocov00}

\section{Extrinsic losses and the sudden approximation}

The importance of extrinsic losses was recognized early, and the landmark
paper here is the discussion by Berglund and Spicer in 1964 of the ''three
step model''. \cite{Berglund64} The three steps are photoexcitation,
transport to the surface, and passage through the surface. In the last two
steps the electrons can loose energy (extrinsic losses). Core electron
photoemission data are often analyzed taking extrinsic losses into account.
Methods to do this have been developed in a long series of papers by
Tougaard and collaborators \cite{Tougaard}. Strong extrinsic losses also
occur in valence electron spectra. In Fig. 1 we show the XPS\ valence band
spectrum from Na metal found by H\"{o}chst, Steiner and H\"{u}fner \cite
{Höchst77}. More details can be found in Refs. \cite{Steiner79}, \cite
{Hüfner95}. It is remarkable that the structure from the plasmon satellites
is as strong as the main band (first peak). Actually the valence electron
spectrum is similar to the core electron spectrum. This similarity is very
close, the full drawn line in Fig. 1 is obtained by convoluting the Na core
spectrum with a valence band spectrum obtained by weighing the partial
densities of states (shown in the inset) with coefficients used as free
parameters (s:p:d=1:5:9). This picture is confirmed by theoretical work. 
\cite{Penn78}, \cite{AHK96}

\begin{figure}
\vspace*{8.cm}
\includegraphics{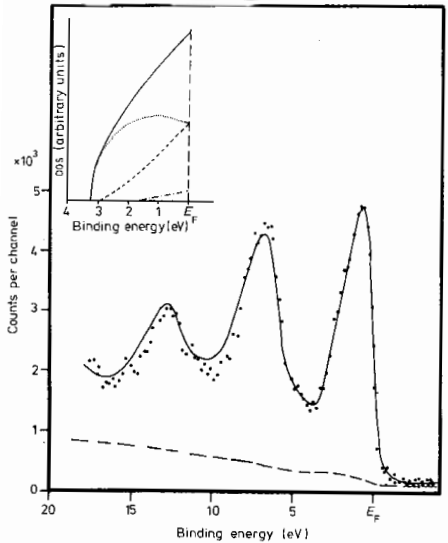}
\caption{XPS valence band spectrum of Na metal. Points are the experimental
data and the full curve is the generated spectrum. The broken curve is the
fitted background. The inset shows the total and partial densities of
states (DOS) which were used to obtain the generated spectrum. Full curve,
total DOS; dotted curve, $s$ partial DOS; broken curve, $p$ partial DOS;
chain curve, $d$ partial DOS.[33]
}
\end{figure}

The plasmon satellites in the core spectrum are easy to understand, the
sudden creation of a core hole changes the potential on the valence
electrons and we have a ''shake-up'' effect. The surprising fact is that the
sudden removal of a valence electron has a very similar shake-up effect, it
is as if the valence electron hole were localized. A quantitative analysis
of core electron satellites showed that both extrinsic and intrinsic losses
contributed comparable amounts. \cite{Pardee75} Since the valence satellites
could be reproduced using the core spectra, we expect that also for valence
electrons both intrinsic and extrinsic losses will contribute.

A theoretical analysis of satellites in simple metals was done by Penn both
for core spectra, \cite{Penn77} and valence spectra, \cite{Penn78}. He
solved a transport equation with the spectral function as source term. The
spectral function for core electrons had been obtained by Langreth. \cite
{Langreth70} For valence electrons Penn obtained a spectral function quite
similar to the core one by doing lowest order perturbation theory using an
effective Hamiltonian based on the Bohm-Pines plasmon theory. Penn's result
clearly demonstrated that both intrinsic and extrinsic effects contributed
to the satellites.

\begin{figure}
\vspace*{6.0cm}
\includegraphics{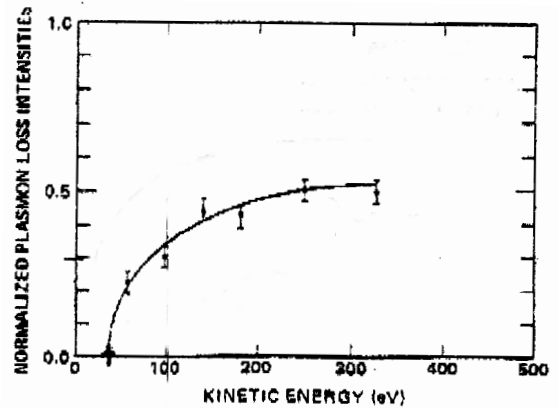}
\caption{Intensity of the first bulk plasmon satellite relative to the main
peak for the $Al\;2p$ core electron spectrum. [45]
}
\end{figure}

As analyzed by Chang and Langreth \cite{Chang72}, \cite{Chang73} there must
be interference between the intrinsic and extrinsic processes since they
lead to the same final state for the system. In particular they discussed
''sp-solids'' and found the interference effects to be strong and fade off
only at kinetic energies in the keV region. A detailed and quantitative
theory of interference effects in a core electron spectrum was developed by
Inglesfield using electron-plasmon coupling functions. \cite{Inglesfield81}, 
\cite{Inglesfield83} Experimental results for the onset of the first plasmon
satellite as function of kinetic energy have been obtained by Flodstr\"{o}m
et al. \cite{Flodström77} (see Fig. 2), and of others (see e. g. Ref. \cite
{Hüfner95}). Their results are well reproduced by Inglesfield's theory. \cite
{Inglesfield83}

A more general approach based on a quasi-boson model, fluctuation potentials
and the GW approximation was developed by Hedin and collaborators. \cite
{Hedin80}, \cite{Bardy85}, \cite{AHK96}, \cite{Hedin99} \ These results also
show that both extrinsic and intrinsic effects are important, and that
interference persists in simple metals to quite high kinetic energies. This
approach has a more rigorous basis and should be quantitatively more
reliable. In particular it handles not only plasmon losses but also the low
energy electron-hole losses which lead to the Mahan-Nozieres-deDominicis
(MND) singular threshold behavior. \cite{Mahan81} We will now discuss
photoemission using this quasi-boson model.

Following earlier work by Hedin et al \cite{Bardy85}, \cite{Hedin98}, \cite
{Hedin99} we use a quasi-boson model Hamiltonian, 
\begin{equation}
H=H_{0}+V,\;H_{0}=H_{syst}+h,  \label{eq:elbosham}
\end{equation}
where 
\[
H_{syst}=\sum_{l}\varepsilon _{l}c_{l}^{\dagger }c_{l}^{{}}+\sum_{q}\omega
_{q}a_{q}^{\dagger }a_{q}^{{}}+\sum_{qll^{\prime }}V_{ll^{\prime
}}^{q}c_{l}^{\dagger }c_{l^{\prime }}^{{}}\left[ a_{q}^{{}}+a_{q}^{\dagger }%
\right] ,
\]
\[
h=\sum_{{\bf k}}\varepsilon _{{\bf k}}^{{}}c_{{\bf k}}^{\dagger }c_{{\bf k}%
}^{{}},\;V=\sum_{q}V^{q}\left[ a_{q}^{{}}+a_{q}^{\dagger }\right]
,\;V^{q}=\sum_{{\bf kk}^{\prime }}V_{{\bf kk}^{\prime }}^{q}c_{{\bf k}%
^{{}}}^{\dagger }c_{{\bf k}^{\prime }}^{{}}
\]
Here indices $"l"$ label states below and $"{\bf k}"$ states above the Fermi
level. Thus $"l"$ can stand for a core electron state or an occupied Bloch
state in the valence band, and $"{\bf k}"$ is a state above the Fermi level
(but possibly below the vacuum level). States below the vacuum level
(occupied and unoccupied) are modified by the presence of the surface and
have exponentially decaying amplitudes outside the solid. States above the
vacuum level have the photoelectron boundary condition, i e they are
time-inverted electron scattering states (LEED\ states) with free electron
energies, $\varepsilon _{{\bf k}}={\bf k}^{2}/2$ and are plane waves at
large distances from the solid. As basis set for the one-electron
wavefunctions we take the eigenfunctions of the Hartree-Fock Hamiltonian. 
\cite{Hedin99} \ We will suppress spin variables. The term $V$ describes the
interaction between the photoelectron and charge density fluctuations (see
below) in the solid, i. e. it is the interaction which causes the extrinsic
losses and also the damping of the photoelectrons. Further $a_{q}$ stands
for quasi-boson type excitations like electron-hole pairs, plasmons etc.,
and $\omega _{q}$ is the quasi-boson energy. The quasi-boson parameters are
related to the dynamically screened potential $W({\bf r},{\bf r}^{\prime
};\omega )$ as described in Appendix A. The potentials $V^{q}\left( {\bf r}%
\right) $ and the wavefunctions can be chosen to be real functions, which we
use in the derivations. The final expressions are valid also with complex
functions.

The model Hamiltonian we have defined clearly can only give an approximate
description of a real system. Thus we obviously do some double counting of
the Coulomb interactions. Despite this the predictions from the model
Hamiltonian are basically sound. For self-energies it correctly describes
the GW approximation, and it gives realistic results for both intrinsic and
extrinsic effects in core electron photoemission. \cite{Hedin99} The
intrinsic satellites in valence electron photoemission for simple metals
come out qualitatively correctly. \cite{AHK96} Further the high energy limit
in photoemission is correctly described. \cite{Hedin98} Clearly a
quasi-boson model cannot handle strongly correlated systems, and we will
discuss these separately.

Returning to Eq. \ref{Psif} we take $\left| N-1,s\right\rangle $ as an
eigenfunction of $H_{syst}$ with eigenvalues $\omega _{s}$, and thus $c_{%
{\bf k}}^{\dagger }\left| N-1,s\right\rangle $ is an eigenfunction of $H_{0}$
with eigenvalue $\varepsilon _{{\bf k}}+\omega _{s}$. Since $E=\varepsilon _{%
{\bf k}}+\omega _{s}$ we can replace $(H-E)$ with $V$. The states $\left|
N-1,s\right\rangle $ have a hole after the emitted photoelectron plus a
number of boson type excitations. For core electron photoemission the bosons
have to be calculated in a potential with a core hole present, while the
initial state $\left| N\right\rangle $ is the vacuum state of bosons with no
core hole. Also in valence electron photoemission, as we have discussed
earlier, the hole after the valence electron has a strong effect on the
quasi-bosons leading to a comparable strength in the satellite. 

For the photoemission amplitude we now have (suppressing the strongly
correlated part in the problem), 
\begin{eqnarray}\label{tau1}
&&\left\langle N-1,s;{\bf k}\left| \Delta \right| N\right\rangle
\\ \nonumber
&=&\sum_{ij}\Delta _{ij}\left\langle N-1,s\left| c_{{\bf k}}\left[ 1+V\frac{1}{%
E-H+i\eta }\right] c_{i}^{\dagger }c_{j}\right| N\right\rangle .
\end{eqnarray}
The physical significance of this expression is quite clear. The operator $%
c_{i}^{\dagger }c_{j}$ lifts one electron from the state $"j"$ into the
state $"i$ $"$. The ''virtual'' electron $"i"$ propagates by $(E-H)^{-1}$
and is then scattered by $V$ into the state $"{\bf k}$ $"$ which is measured
by the detector. The state $"i"$ must be a photoelectron state, which we
denote by $\left| {\bf k}^{\prime }\right\rangle $. Since we never have more
than one photoelectron we can work with a product representation, 
\[
c_{{\bf k}}^{\dagger }\left| N-1,s\right\rangle =\left| N-1,s\right\rangle
\left| {\bf k}\right\rangle ,\;c_{{\bf k}^{\prime }}^{\dagger }\left|
N\right\rangle =\left| N\right\rangle \left| {\bf k}^{\prime }\right\rangle
. 
\]
We can now rewrite Eq. \ref{tau1} as 
\begin{eqnarray}\label{tau2}
& &\left\langle N-1,s;{\bf k}\left| \Delta \right| N\right\rangle
\\ \nonumber &=&\sum_{j{\bf k%
}^{\prime }}\left\langle {\bf k}\left| \left\langle N-1,s\left| \left[ 1+V%
\frac{1}{E-H+i\eta }\right] c_{j}\right| N\right\rangle \right| {\bf k}%
^{\prime }\right\rangle
\\ \nonumber & & \ \ \ \times
\left\langle {\bf k}^{\prime }\left| \Delta \right|
j\right\rangle \\ \nonumber &=&
\sum_{j}\left\langle {\bf k}\left| \left\langle N-1,s\left| \left[ 1+V\frac{%
1}{E-H+i\eta }\right] c_{j}\right| N\right\rangle \Delta \right|
j\right\rangle ,
\end{eqnarray}
noting that $\sum_{{\bf k}^{\prime }}\left| {\bf k}^{\prime }\right\rangle
\left\langle {\bf k}^{\prime }\right| $ acts as a deltafunction.

The potential $V$ in Eq. \ref{tau2} gives the two effects of a loss
potential, the energy loss scattering and the related reduction in the
intensity of the no-loss peak. The latter comes through a partial summation
over the no-loss intermediate state, which gives a damped photoemission
wavefunction. Formally this summation can be done by use of the Feshbach
projection operator technique. The result is (see e g Ref. \cite{Hedin98}) 
\begin{eqnarray}\label{tau3}
&&\left\langle N-1,s;{\bf k}\left| \Delta \right| N\right\rangle=
\\ \nonumber
&&\sum_{j}\left\langle \widetilde{{\bf k}}\left| \left\langle N-1,s\left| %
\left[ 1+V\frac{1}{E-QHQ+i\eta }\right] c_{j}\right| N\right\rangle \Delta
\right| j\right\rangle ,  
\end{eqnarray}
where $\left| \widetilde{{\bf k}}\right\rangle $ is the wavefunction
discussed in connection with Eq. \ref{heff}, and $Q$ is the projection
operator $Q=1-P$, with $P=\left| N-1,s\right\rangle \left\langle
N-1,s\right| $. In principle $\left| \widetilde{{\bf k}}\right\rangle $ and
the self-energy, 
\[
\Sigma =PVQ\frac{1}{E-QHQ+i\eta }QVP,
\]
depend on the excitation state of the target, $"s"$. To lowest order we have 
\cite{FH89} 
\begin{equation}
\Sigma \left( {\bf r},{\bf r}^{\prime };\omega \right) =\sum_{q}V^{q}\left( 
{\bf r}\right) \frac{1}{\omega -h-\omega _{q}+i\eta }V^{q}\left( {\bf r}%
^{\prime }\right) ,  \label{Sigma}
\end{equation}
provided the quasi-bosons in the state $"s"$ are extended and thus a finite
number of them have a negligible influence. Also when we go to higher orders
the result remains independent on the index $"s"$. The imaginary part of $%
\Sigma $ in Eq. \ref{Sigma} agrees with the imaginary part in the GW
approximation for the self-energy \cite{FH89}, while the real parts differ
slightly.

To estimate the extrinsic and intrinsic losses we expand Eq. \ref{tau3} to
lowest non-trivial order in $V^{q}$ (see Appendix B). To lowest order in $V$
there is one contribution from $V\left( E-H_{0}\right) ^{-1}$ (extrinsic
loss) and one from $\left\langle N-1,s\left| c_{j}\right| N\right\rangle $
(intrinsic loss described by the spectral function). In both cases $"s"$ is
a state with one quasiboson $\left( q\right) $ and one hole. The result is
given in Eqs. \ref{Jk3} and \ref{tau4}. The selfenergy $\Sigma $ in the
denominator does not appear in the simplest lowest order perturbation theory
but is obtained in a systematic expansion. \cite{Bardy85}, \cite{FH89}

We first discuss valence electron photoemission when the index $l$ is a
vector index ${\bf l}$. From the discussion at the end of Sect II we see
that the matrix element $\left\langle \widetilde{{\bf k}}\left| \Delta
\right| {\bf l}\right\rangle $ is peaked around ${\bf l=l}^{0}$. The no-loss
contribution to $J_{{\bf k}}\left( \omega \right) $ is 
\begin{eqnarray*}
J_{{\bf k}}^{0}\left( \omega \right)&=&\sum_{{\bf l}}\left| \left\langle
\widetilde{{\bf k}}\left| \Delta \right| {\bf l}\right\rangle \right|
^{2}\delta \left( \omega -\varepsilon _{{\bf k}}+\varepsilon _{{\bf l}%
}\right) \\ \nonumber
&=&\frac{1}{c^{2}}\int \frac{\left| \left\langle \widetilde{{\bf k}}%
\left| \Delta \right| {\bf l}^{0}\right\rangle _{0}\right| ^{2}}{\left(
l_{z}^0-l_{z}\right) ^{2}+\left( k_{f}^{I}\right) ^{2}}\delta
\left( \omega -\varepsilon _{{\bf k}}+\varepsilon _{{\bf l}}\right) dl_{z}
\\ \nonumber
&=&\frac{1}{c^{2}}\frac{1}{\left| \partial \varepsilon _{{\bf l}}/\partial
l_{z}\right| }\frac{\left| \left\langle \widetilde{{\bf k}}\left| \Delta
\right| {\bf l}^{0}\right\rangle _{0}\right| ^{2}}{\left( l%
_{z}^0-l_{z}^{1}\right) ^{2}+\left( k_{f}^{I}\right) ^{2}},
\end{eqnarray*}
which is the QP particle result in Eq. \ref{Jk5} in the limit when $\Gamma
_{i}$ (and $\Sigma _{{\bf l}}$) are infinitesimal.

For the loss contribution to $J_{{\bf k}}\left( \omega \right) $ we first
consider only {\it the intrinsic part} (c.f. Eqs \ref{Jk} and \ref{tau4}), 
\begin{equation}
J_{{\bf k}}^{1}\left( \omega \right) =\sum_{q{\bf l}}\left| \sum_{{\bf l}%
^{\prime }}\frac{-\left\langle \widetilde{{\bf k}}\left| \Delta \right| {\bf %
l}^{\prime }\right\rangle V_{{\bf l}^{\prime }{\bf l}}^{q}}{\omega
_{q}-\varepsilon _{{\bf l}}+\varepsilon _{{\bf l}^{\prime }}}\right|
^{2}\delta \left( \omega -\omega _{q}-\varepsilon _{{\bf k}}+\varepsilon _{%
{\bf l}}\right) .  \label{Jkintr}
\end{equation}
The quadratic expression in the fluctuation potentials can be written in
terms of the imaginary part of the selfenergy 
\[
\mathop{\rm Im}%
\Sigma _{{\bf l}^{\prime }{\bf l}^{\prime \prime }}\left( \omega \right)
=\pi \sum_{q{\bf l}}V_{{\bf l}^{\prime }{\bf l}}^{q}V_{{\bf l}^{\prime
\prime }{\bf l}}^{q}\delta \left( \omega +\omega _{q}-\varepsilon _{{\bf l}%
}\right) ,
\]
provided $\omega <\mu $, (see Eq. 19 in Ref. \cite{Hedin99}). We can express 
$%
\mathop{\rm Im}%
\Sigma \left( \omega \right) $ in terms of $%
\mathop{\rm Im}%
W\left( \omega \right) $ as is clear from Eq. \ref{ImW}.This gives 
\[
J_{{\bf k}}^{1}\left( \omega \right) =\frac{1}{\pi }\sum_{{\bf l}^{\prime }%
{\bf l}^{\prime \prime }}\frac{\left\langle \widetilde{{\bf k}}\left| \Delta
\right| {\bf l}^{\prime }\right\rangle \left\langle \widetilde{{\bf k}}%
\left| \Delta \right| {\bf l}^{\prime \prime }\right\rangle 
\mathop{\rm Im}%
\Sigma _{{\bf l}^{\prime }{\bf l}^{\prime \prime }}\left( \varepsilon _{{\bf %
k}}-\omega \right) }{\left( \omega -\varepsilon _{{\bf k}}+\varepsilon _{%
{\bf l}^{\prime }}\right) \left( \omega -\varepsilon _{{\bf k}}+\varepsilon
_{{\bf l}^{\prime \prime }}\right) }.
\]
The condition $\omega <\mu $ for $%
\mathop{\rm Im}%
\Sigma \left( \omega \right) $ translates to $\varepsilon _{{\bf k}}-\omega
<\mu $ in the above equation. This means that $\omega >\varepsilon _{{\bf k}%
}-\mu $, where $\omega $ is the photon energy, which clearly is satisfied.
Since $%
\mathop{\rm Im}%
\Sigma _{{\bf l}^{\prime }{\bf l}^{\prime \prime }}\left( \omega \right) $
is diagonal to a good approximation (Ref. \cite{Hedin99}) we take 
\begin{equation}
J_{{\bf k}}^{1}\left( \omega \right) =\frac{1}{\pi }\sum_{{\bf l}^{\prime }}%
\frac{\left| \left\langle \widetilde{{\bf k}}\left| \Delta \right| {\bf l}%
^{\prime }\right\rangle \right| ^{2}%
\mathop{\rm Im}%
\Sigma _{{\bf l}^{\prime }}\left( \varepsilon _{{\bf k}}-\omega \right) }{%
\left( \omega -\varepsilon _{{\bf k}}+\varepsilon _{{\bf l}^{\prime
}}\right) ^{2}}.  \label{Jk1}
\end{equation}

We now make the following Ansatz for $J_{{\bf k}}\left( \omega \right) $%
\begin{eqnarray}\label{Jexp}
J_{{\bf k}}\left( \omega \right)&=&\sum_{{\bf l}}\left| \left\langle 
\widetilde{{\bf k}}\left| \Delta \right| {\bf l}\right\rangle \right|
^{2}\int_{-\infty }^{\infty }\frac{dt}{2\pi }e^{i\left( \omega -\varepsilon
_{{\bf k}}+E_{{\bf l}}\right) t}
\\ \nonumber
&\times&\exp \left( \frac{1}{\pi }\int \frac{%
\mathop{\rm Im}%
\Sigma _{{\bf l}}\left( \varepsilon _{{\bf l}}-\omega ^{\prime }\right)
\left( e^{-i\omega ^{\prime }t}-1\right) d\omega ^{\prime }}{\left( \omega
^{\prime }\right) ^{2}}\right) . 
\end{eqnarray}
The integrand in the exponent is the function $S_{{\bf l}}\left( -t\right) $
discussed in Refs \cite{Hedin80} and \cite{AHK96}, 
\[
S_{{\bf l}}\left( t\right) =\frac{1}{\pi }\int \frac{%
\mathop{\rm Im}%
\Sigma _{{\bf l}}\left( \varepsilon _{{\bf l}}-\omega \right) \left(
e^{i\omega t}-1\right) d\omega }{\omega ^{2}},
\]
which for large $t$ is 
\[
S_{{\bf l}}\left( t\right) =-\Gamma _{{\bf l}}\left| t\right| +i\alpha _{%
{\bf l}}sgn\left( t\right) -n_{{\bf l}}.
\]
The quasiparticle behavior is set by the large $t$ limit. Neglecting the
asymmetry index $\alpha _{{\bf l}}$ we have 
\[
J_{{\bf k}}^{QP}\left( \omega \right) =\frac{1}{\pi }\sum_{{\bf l}}\left|
\left\langle \widetilde{{\bf k}}\left| \Delta \right| {\bf l}\right\rangle
\right| ^{2}e^{-n_{{\bf l}}}\frac{\Gamma _{{\bf l}}}{\left( \omega
-\varepsilon _{{\bf k}}+E_{{\bf l}}\right) ^{2}+\Gamma _{{\bf l}}^{2}},
\]
which agrees with Eqs \ref{Jk4} and \ref{Al} since the factor $\exp \left(
-n_{{\bf l}}\right) $ equals $Z_{l}$ to leading order. \cite{AHK96} The
Ansatz to leading order also reproduces the intrinsic loss $J_{{\bf k}%
}^{1}\left( \omega \right) $ in Eq. \ref{Jk1} apart from a normalization
factor. The Ansatz further agrees with the result for intrinsic losses
obtained earlier. \cite{AHK96} To have the intrinsic core electron spectrum
we just replace the index ${\bf l}$ by $c$ in Eq. \ref{Jexp}, and we have no
summation.

For core electrons we can obtain {\it both the intrinsic and extrinsic
contributions} \cite{Inglesfield83}, a detailed derivation is given in
Appendix A\ in Ref. \cite{HL01}. We found that Eq. \ref{tau4} can be written 
\begin{eqnarray*}
\tau _{qc}\left( {\bf k}\right)&=&-\frac{V^{q}\left( {\bf r}_{c}\right) }{%
\omega _{q}}\left\langle \widetilde{k}_{z},{\bf K}\left| \Delta \right|
c\right\rangle
\\ \nonumber
&+&\frac{1}{i\kappa }\int_{-\infty }^{z_{0}}dz\psi _{\widetilde{%
k}_{z}}^{>}\left( z\right) V\left( q_{z},{\bf Q},z\right) \psi _{\kappa
}^{<}\left( z\right) \\ \nonumber
& & \ \ \ \ \ \ \ \ \ \ \ \ \ \times\left\langle \kappa ,{\bf K-Q}\left| \Delta
\right|
c\right\rangle ,
\end{eqnarray*}
provided we make a number of simplifying approximations. We have e. g. taken
the crystal as translationally invariant parallel to the surface. ${\bf K}$
and ${\bf Q}$ are the photoelectron and fluctuation potential momenta
parallel to the surface, and ${\bf r}_{c}$ is the position of the atom from
which the core electron is excited. Further $V\left( q_{z},{\bf Q},z\right) $
is the fluctuation potential $V^{q}$, and $\psi _{\widetilde{k}%
_{z}}^{>}\left( z\right) $ and $\psi _{\kappa }^{<}\left( z\right) $ are
damped Blochfunctions, $\psi _{\kappa }^{<}$ decreasing towards the surface,
and $\psi _{\kappa }^{>}$ decreasing towards the inner of the crystal. The
momenta $\widetilde{k}_{z}$ and $\kappa $ are equal when the quasi-boson
energy and momenta are zero. We also take the dipole matrix elements as
equal, and have 
\begin{eqnarray}\label{tau5}
& &\tau _{qc}\left( {\bf k}\right)
\\ \nonumber
&=&\left\langle \widetilde{k}_{z},{\bf K}%
\left| \Delta \right| c\right\rangle
\\ \nonumber
&\times&\left[ -\frac{V^{q}\left( {\bf r}%
_{c}\right) }{\omega _{q}}+\frac{1}{i\kappa }\int_{-\infty }^{z_{0}}dz\psi _{%
\widetilde{k}_{z}}^{>}\left( z\right) V\left( q_{z},{\bf Q},z\right) \psi
_{\kappa }^{<}\left( z\right) \right] ,  
\end{eqnarray}
We can now obtain the result with both intrinsic and extrinsic contributions
in the core electron case from the analysis of the intrinsic case by
following the derivation from Eq. \ref{Jkintr} to Eq. \ref{Jexp} and
replacing $-V^{q}\left( {\bf r}_{c}\right) /\omega _{q}$ by the full square
parenthesis in Eq. \ref{tau5}. The result is given in Eq. \ref{Jkcore}. The
expression in Eq. \ref{tau4} for the extrinsic part for valence electrons is
more involved to simplify. From the similar structure of the expressions for
the core and valence electron cases we can however expect a similar
qualitative behavior.

\section{The use of the electron energy loss function to estimate losses in
photoemission}

It is is a heavy computational task to do a priori calculations of the
fluctuation potentials that are needed to evaluate extrinsic and intrinsic
losses. Even when we limit ourselves (as we have done here) to expressions
quadratic in the potentials, which can be described by $%
\mathop{\rm Im}%
W\left( {\bf r},{\bf r}^{\prime };\omega \right) $, the task is heavy since
it is important to allow for the behavior of $W$ at the surface. To avoid
these difficulties, awaiting advances in computational techniques, we will
develop approximations for $W$ based on knowledge of the experimental energy
loss function $-%
\mathop{\rm Im}%
\varepsilon ^{-1}\left( {\bf q},\omega \right) $. \cite{HL01} This is done
in three steps. First we connect the bulk expression for $%
\mathop{\rm Im}%
W\left( {\bf r},{\bf r}^{\prime };\omega \right) $ to $%
\mathop{\rm Im}%
\varepsilon ^{-1}\left( {\bf q},\omega \right) $, and then we modify $%
\mathop{\rm Im}%
W\left( {\bf r},{\bf r}^{\prime };\omega \right) $ to account for the
surface. Unfortunately not much is known about energy loss at finite
momenta, and one hence usually also have to make the third step of
extrapolating the experimental $%
\mathop{\rm Im}%
\varepsilon ^{-1}\left( {\bf q=0},\omega \right) $ values to finite ${\bf q}$
values.

For a simple metal or a valence semiconductor, an ''$sp$-system'' \cite
{Hedin99}, we take $\varepsilon ^{-1}$ as diagonal in ${\bf q}$ space.\ We
then have 
\[
\mathop{\rm Im}%
W^{bulk}\left( {\bf r},{\bf r}^{\prime };\omega \right) =\int e^{i{\bf q}%
\left( {\bf r}-{\bf r}^{\prime }\right) }v\left( {\bf q}\right) 
\mathop{\rm Im}%
\varepsilon ^{-1}\left( {\bf q},\omega \right) d{\bf q.} 
\]
The fluctuation potentials for a bounded jellium have been discussed in Ref. 
\cite{Hedin98}. The results for photoemission turned out to be fairly
insensitive to the precise form of the potentials as long as they went to
zero at the surface. A reasonable choice is the Inglesfield one \cite
{Inglesfield83} 
$$
V^{q}\left( z\right)=F\left( z;q_{z},Q\right) V_{0}^{q},
$$
\begin{equation}\label{Vq}
F\left(z;q_{z},Q\right)
=\left[ \cos \left( q_{z}z+\phi _{q}\right) -\cos \phi
_{q}e^{-Qz}\right] \theta \left( z\right) ,
\end{equation}
which is zero at the surface ( $z=0$) and rapidly turns into a sine wave.
The phase $\phi _{q}$ is a certain function of $q_{z}$ and $Q$. This leads
us to replace the bulk expression 
\[
\mathop{\rm Im}%
W\left( z,z^{\prime };{\bf Q},\omega \right) \simeq
\sum_{q_{z}}e^{iq_{z}\left( z-z^{\prime }\right) }%
\mathop{\rm Im}%
W\left( q_{z};{\bf Q},\omega \right) 
\]
by 
\begin{eqnarray*}
&&\mathop{\rm Im}%
W\left( z,z^{\prime };{\bf Q},\omega \right)
\\ \nonumber &\simeq& \sum_{q_{z}}F\left(
z;q_{z},Q\right) F\left( z^{\prime };q_{z},Q\right)
\mathop{\rm Im}%
W\left( q_{z};{\bf Q},\omega \right) .
\end{eqnarray*}

For correlated quasi 2D systems we are far from translational invariance in
the $z$ direction. Suppose we still have translational invariance parallel
to the planes. We write the response between the induced charge density and
the total (external plus induced) potential as $\chi ^{0}$, $\rho
^{ind}=\chi ^{0}V^{tot}$, and approximate 
\begin{eqnarray}\label{chi0}
\chi ^{0}\left( z,z^{\prime };{\bf Q}\right)
&=&\chi _{b}^{0}\left(
z-z^{\prime };{\bf Q}\right) \\ \nonumber 
&+&\sum_{mn}w\left( z-cm-d_{n}\right) w\left(
z^{\prime }-cm-d_{n}\right) \widetilde{\chi }_{2}^{0}\left( {\bf Q}\right) .
\end{eqnarray}
Here $w\left( z\right) =\left| \phi _{0}\left( z\right) \right| ^{2}$, where 
$\phi _{0}\left( z\right) $ is the ground state wave function for electrons
localized in a 2D plane. The length of the unit cell in the $z$ direction is 
$c$, and $d_{n}$ gives the position of the layers inside a unit cell (when
there are more than one layer). We thus write the response to the total
potential as one part coming from extended 3D excitations plus one part from
electrons which are localized in the $z$ direction and only move in the
planes. The variable ${\bf Q}$ comes from the Fourier transform of ${\bf R}-%
{\bf R}^{\prime }$, the coordinate distance in the plane. The susceptibility
relating the induced charge density to the external potential $\chi $, $\rho
^{ind}=\chi V^{ext}$ is related to $\chi ^{0}$ by $\chi =\chi ^{0}+\chi
^{0}v\chi $, and the inverse dielectric function is $\varepsilon
^{-1}=1+v\chi $. Eq. \ref{chi0} leads to a 3D bulk contribution to $%
\mathop{\rm Im}%
W\left( z,z^{\prime }\right) $, plus a contribution $v\chi _{2}v$, where the
susceptibility $\chi _{2}$ is calculated as if we only had the 2D planes and
then replacing the bare Coulomb potentials in $\chi _{2}$ with dynamically
screened potentials from the embedding 3D electrons. We note that the bulk
screened potential can be anisotropic since $\chi _{b}^{0}\left( q_{z},{\bf Q%
}\right) $ can depend on both $q_{z}$ and ${\bf Q}$, and not only on $%
q_{z}^{2}+{\bf Q}^{2}$. If there is only one layer in the unit cell there is
an explicit solution of the integral equation $\chi _{2}=\chi _{2}^{0}+\chi
_{2}^{0}v\chi _{2}^{{}}$, 
\[
\chi _{2}^{{}}\left( q_{z},q_{z}^{\prime }\right) =\frac{1}{c}w\left(
q_{z}\right) w\left( q_{z}^{\prime }\right) \widetilde{\chi }_{2}^{{}}\left(
q_{z}\right) .
\]
The variables ${\bf Q}$ and $\omega $ are suppressed, $w\left( q_{z}\right) $
is the Fourier transform of $w\left( z\right) $ and $\widetilde{\chi }%
_{2}^{{}}\left( q_{z}\right) $ is a periodic function of $q_{z}$. Clearly $%
w\left( 0\right) =1$, and often it is reasonable to take $w\left(
q_{z}\right) =1$ for all $q_{z}$. Due to the periodicity $q_{z}$ and $%
q_{z}^{\prime }$ must differ by a reciprocal lattice vector. This means that
if we know the diagonal part of $\chi _{2}^{{}}\left( q_{z},q_{z}^{\prime
}\right) $ (from the loss function) we also know the full $\chi
_{2}^{{}}\left( q_{z},q_{z}^{\prime }\right) $. With several layers in the
cell we have 
\[
\chi _{2}^{{}}\left( q_{z}^{{}},q_{z}^{\prime }\right) =\frac{1}{c}w\left(
q_{z}^{{}}\right) w\left( q_{z}^{\prime }\right) \sum_{nn^{\prime }}%
\widetilde{\chi }_{nn^{\prime }}\left( q_{z}\right)
e^{iq_{z}^{{}}d_{n^{{}}}-iq_{z}^{\prime }d_{n^{\prime }}}.
\]
Here $\widetilde{\chi }_{nn^{\prime }}\left( q_{z}\right) $ is periodic, but
the exponential factor is not. The diagonal part of $\chi _{2}^{{}}\left(
q_{z}^{{}},q_{z}^{\prime }\right) $ contains a sum of the matrix elements $%
\widetilde{\chi }_{nn^{\prime }}\left( q_{z}\right) $ and thus the loss
function cannot give us all of them. The diagonal elements of $\widetilde{%
\chi }_{nn^{\prime }}\left( q_{z}\right) $ are equal, and if we neglect the
non-diagonal elements (the coupling between the layers in a cell), again we
have only one unknown quantity, which we can get from the loss function.
Estimates for Bi2212 show that the nondiagonal elements are smaller than the
diagonal ones \cite{HL01}, but still it is a fairly crude approximation to
neglect them. In lack of other information than the loss function, we
however feel forced to do so.

\begin{figure}
\vspace*{6.7cm}
\includegraphics{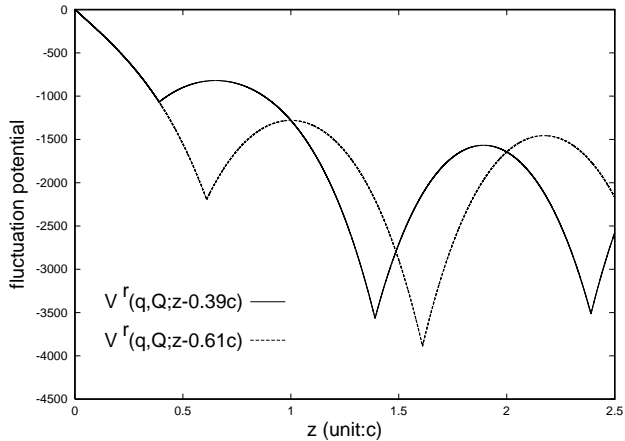}
\caption{The fluctuation potentials $V^{r}\left( z-z_{1}\right)
$ and $V^{r}\left( z-z_{2}\right) $ where $z_{1}$ and $z_{2}$ are the
positions of the first and second CuO layers in Bi2212. Note that in both
cases the potentials are zero at the surface, $z=0$.
}
\end{figure}

To obtain the 2D contribution to $W\left( z,z^{\prime }\right) $ we have to
sum 
\[
\sum_{q_{z}^{{}}q_{z}^{\prime }}e^{-iq_{z}z}e^{iq_{z}^{\prime }z^{\prime
}}v\left( q_{z}\right) \chi _{2}^{{}}\left( q_{z},q_{z}^{\prime }\right)
v\left( q_{z}^{\prime }\right) , 
\]
where the variables ${\bf Q}$ and $\omega $ still are suppressed. We split
the summation over $q_{z}$ into a sum (integral) over the Brillouin zone and
a sum over reciprocal vectors $G$, 
\begin{eqnarray*}
&&W\left( z,z^{\prime }\right) -W^{3D}\left( z-z^{\prime }\right)
\\ \nonumber &\simeq&
\sum_{n}\int_{-\pi /c}^{\pi /c}V\left( z-d_{n},q_{z}\right) V\left(
z^{\prime }-d_{n},q_{z}\right) \widetilde{\chi }_{1,1}\left( q_{z}\right)
dq_{z},
\end{eqnarray*}
where 
$$
V\left( z,q_{z}\right) =e^{-iq_{z}z}V^{p}\left( z,q_{z}\right),
$$
\begin{equation}
V^{p}\left( z,q_{z}\right) =\sum_{G}e^{-iGz}v\left( q_{z}+G\right)
w\left( q_{z}+G\right) .  \label{V(z)}
\end{equation}
The spacial dependence of the fluctuation potential is thus that of a Bloch
wave, a plane wave times a periodic function $V^{p}$. $V^{p}$ is a
well-known function that can be obtained explicitly if we take $w\left(
q_{z}\right) =1$. We define real functions 
\[
V^{r}\left( z,q_{z}\right) =%
\mathop{\rm Re}%
\left[ e^{-iq_{z}z+i\phi \left( q_{z},z_{0}\right) }V^{p}\left(
z,q_{z}\right) \right] , 
\]
and choose the phaseshift $\phi \left( q_{z},z_{0}\right) $ to have $%
V^{r}\left( z,q_{z}\right) =0$ at the surface $z=z_{0}$. This procedure is
like we would have taken $F\left( z;q_{z}\right) =\sin [q_{z}\left(
z-z_{0}\right) ]$ which is not unreasonable. In summary this gives us 
\begin{eqnarray*}
&&\mathop{\rm Im}%
W^{2D}\left( z,z^{\prime }\right) \\ \nonumber
&\simeq& \sum_{n}\int_{-\pi /c}^{\pi
/c}V^{r}\left( z-d_{n},q_{z}\right) V^{r}\left( z^{\prime
}-d_{n},q_{z}\right)
\mathop{\rm Im}%
\widetilde{\chi }_{1,1}\left( q_{z}\right) dq_{z},
\end{eqnarray*}
where $%
\mathop{\rm Im}%
\widetilde{\chi }_{1,1}\left( q_{z}\right) $ can be obtained from the loss
function. We should have replaced $v\left( q_{z}\right) $ by a screened
potential $W\left( q_{z}\right) $ everywhere, but since we have the same
replacements in the expression for loss function, this is of no consequence.
The coupling functions $V^{r}\left( z\right) $ are quite different from the
functions $V^{q}\left( z\right) $ in Eq. \ref{Vq} for the 3D case. As shown
in Fig. 3 they look more like a superposition of surface plasmon functions
centered at the different 2D layers then a sine wave.

The third step of extrapolating to finite momenta also involves considerable
uncertainty. It is however important to do some estimates of the effects of
dispersion since we know that for an electron gas it makes a difference of a
factor of two for the mean free path. For an electron gas the plasmon pole
model is fairly good, also when a simple quadratic dispersion is used 
\[
\mathop{\rm Im}%
\frac{-1}{\varepsilon \left( q,\omega \right) }\simeq \delta \left( \omega
-\omega _{q}\right) ,\;\omega _{q}=\omega _{p}+q^{2}/2. 
\]
Often a one mode description is not good enough. One can then use a sum of
Drude functions, which further include damping of the quasi-boson modes, 
$$
\mathop{\rm Im}%
\frac{-1}{\varepsilon \left( q,\omega \right) }=\sum_{n}c_{n}^{0}\frac{%
\omega \Gamma _{n}\left( q\right) }{\left( \omega ^{2}-\omega _{n}^{2}\left(
q\right) \right) ^{2}+\omega ^{2}\Gamma _{n}^{2}\left( q\right) },
$$
\begin{equation}\label{lossfunc}
\omega
_{n}\left( q\right) =\omega _{n}^{0}+q^{2}/2,
\end{equation}
as suggested by Ritchie and Howie \cite{Ritchie77} and others. \cite
{Tougaard84} \ The dispersion of $\omega _{n}\left( q\right) $ is somewhat
ad hoc, however for large $q$ it has the correct form, following the Bethe
ridge. Also for $\Gamma _{n}\left( q\right) $ some dispersion could be used.
The coefficients $c_{n}^{0},\;\omega _{n}^{0}$, and $\Gamma _{n}\left(
0\right) $ can be determined by fitting to the energy loss function $-%
\mathop{\rm Im}%
\varepsilon ^{-1}\left( q=0,\omega \right) $.

\section{Interference between intrinsic and extrinsic losses}

\subsection{Localized strongly correlated systems}

There is a fundamental difference between photoemission from solids and from
localized systems like atoms or molecules. The interaction between the
photoelectron and the target always goes to zero at high kinetic energies.
In solids this decreasing interaction is compensated by the longer distance
the electron travels, due to its longer mean free path, before it leaves the
solid (assuming that the light penetrates much deeper into the solid than
the mean free path). Thus while at high energy the photoelectron coupling
can be neglected for a localized system, it never becomes negligible for a
solid.

An extensive literature has appeared on near-threshold behavior of
satellites in atomic spectra, with a landmark paper from 1965 by Carlson and
Krause \cite{Carlson65} where they observed the characteristic shake-up
behavior for the first time. They showed that the intensity of the shake-up
satellite in K-shell excitation of the Ne atom rouse slowly from zero at
threshold to its sudden approximation limit at some $200\;eV$ higher energy.
The data were plotted in units of the ''excitation'' energy $E_{0}$ of the
satellite final state relative to the main line ($E_{0}=50\;eV$ in the Ne
case). The onset of satellite intensity was studied particularly in the
80's when good synchrotron sources had become available. St\"{o}hr,
Jaeger and Rehr in 1983 \cite{Stöhr83} made a systematic study of the energy
dependence of a core level shake-up feature (the K spectrum of $N_{2}$ on $%
Ni\left( 100\right) $) with a much smaller value of $E_{0}$ (5 eV). Also
here the approach to the sudden limit was found to occur within a few unit
of $E_{0}$. Reviews of the atomic work on threshold satellites have been
given by Becker and Shirley. \cite{Shirley87}, \cite{Becker90} The picture
that evolved was complex with rapid variations in satellite intensities near
threshold. Much of the complexity is however related to Rydberg levels and
post collision effects, which have less relevance in solids.

The satellite turn-on was recently studied in detail for a numerically
exactly soluble model that should be generic for a class of strongly
correlated solid state systems.\cite{Lee99} The model has a charge transfer
excitation, and is relevant for transition metal halides, rare earth
compounds, chemisorption systems, and high $T_{c}$ compounds. It is modelled
by three electron levels, one core level and two outer levels. When the core
hole (3s) is created, the more localized outer level (d) is pulled below the
less localized level (L). The spectrum has a leading peak corresponding to a
charge transfer between L and d (''{\it shake-down}''), and a satellite
corresponding to no charge transfer. The ''{\it conjugate shake-up}''
mechanism (optical transition between two localized levels (s and d) plus
shake-up from a localized level (L) to the continuum) analyzed by Thomas 
\cite{Thomas84} is not relevant in our case since an s-d transition is
optically forbidden. The crucial feature of the model is the dynamic Coulomb
interaction between the photoelectron and the d and L levels, i. e. the
interaction that causes extrinsic losses. The calculations were done with
parameters relevant to different copper-dihalide compounds ($%
CuBr_{2},\;CuCl_{2},CuF_{2}$), and showed a rapid approach (see Fig. 4) to
the sudden limit within 5-10 eV, which is in accord with the St\"{o}hr et al
results. To the extent that the localized system described by this model can
be regarded as decoupled from the rest of the (3D) solid, this means that
above say 10 eV extrinsic losses and interference effects are associated
only with 3D extended excitations.

\begin{figure}
\vspace*{6.cm}
\includegraphics{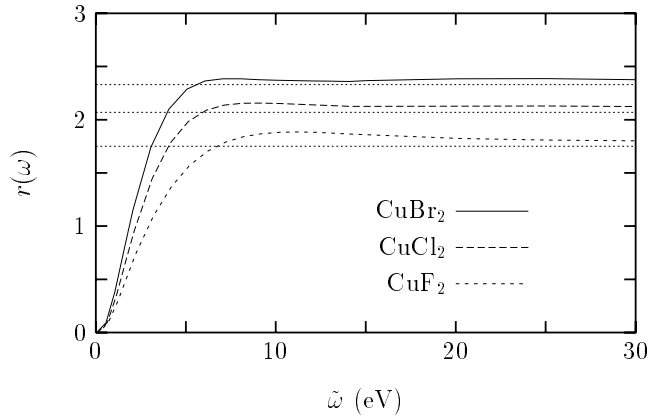}
\caption{The ratio $r\left( \omega \right) $ between the
satellite and the main peak photocurrents for some divalent copper
compounds. The energy is measured from the threshold for satellite onset,
and the horizontal lines give the sudden limit.
}
\end{figure}

With an exactly solvable model some detailed mechanisms for the satellite
turn-on could be studied. Since the transition takes place at such low
energies it turned out that the variation of the dipole matrix element gave
effects comparable to that of the dynamic photoelectron coupling. When the
dipole matrix effect dominates (intrinsic case), the onset is determined by $%
E_{0}$ as long as $E_{0}$ is smaller or comparable to the characteristic
energy $E_{d}=1/R_{d}^{2}$ where $R_{d}$ gives the length scale of the
dipole matrix element. For $E_{0}$ larger than $E_{d}$ it is instead $E_{d}$
which gives the characteristic onset energy. The strength of the satellite
relative to that of the main line obtained with the full model divided by
the result when only the matrix element was considered, showed curves
starting at 1 at zero energy which could have sizable humps (extrinsic
effects) at an energy of order $E_{d}$. Attempts to describe these effects
by a semiclassical approximation were not successful, and no simple picture
in time space evolved. Lowest order perturbation theory on the other hand
succeeded in giving some qualitative understanding. One conclusion is that $%
E_{0}$ may not always be the correct scaling parameter for satellite onset.

\subsection{Weakly correlated solids}

The transition from the adiabatic to the sudden limit in core electron
photoemission for weakly correlated systems was recently studied. \cite
{Hedin98} The theory is fully quantum mechanical and includes the
interference between intrinsic and extrinsic losses (QM). It is a general
theory for weakly correlated \ systems and describes not only plasmon losses
but losses to density fluctuations in general like those from electron-hole
excitations. The actual calculations were done for an atom embedded in a
semi-infinite jellium. For such a simple model explicit and detailed results
could be obtained. The limit expression when the interference term in Eq. 
\ref{tau5} was neglected (BS), gives essentially the same approximation as
discussed by Berglund and Spicer\cite{Berglund64} and used by Penn. \cite
{Penn77} Comparison was also made with results from the semi-classical
theory (SC). In this theory the photoelectron is taken to travel with a
steady speed on a straight trajectory, and the photoelectron losses are
identified with the excitations caused in the solid by the external
perturbation from the moving classical photoelectron. In Sect. III we
presented general results, partly based on Ref. \cite{Hedin98}.

\begin{figure}
\vspace*{6.7cm}
\includegraphics{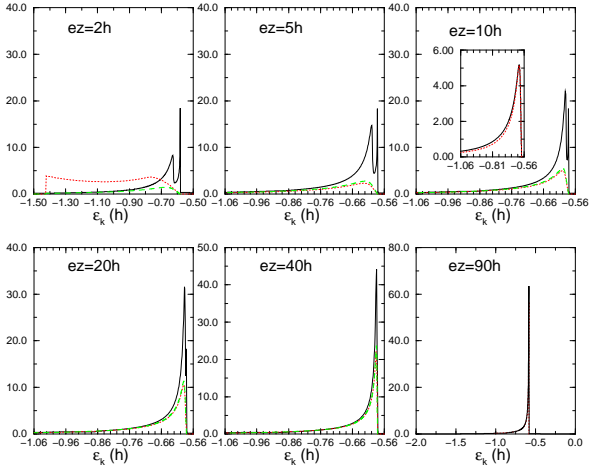}
\caption{Satellite part of the photoelectron
spectrum for Al ($r_{s}=2.07$) from lowest order perturbation theory showing
the contribution from bulk plasmons as function of photoelectron energy (the
zero of the energy scale is at the quasi-particle energy), for different
photon energies. We show results for three different
approximations, fully quantum mechanical (QM) from Eq. \ref{gQM} (dotted
curves); neglecting interference (BS) from Eq. \ref{gBS} (solid curves); and
semiclassical (SC) from Eq. \ref{gSC} (dashed curves). The inset in the $%
ez=10$ curve shows results with two different polarization potentials
(see Ref.[21]).
}
\end{figure}

The photocurrent for the core electron case is 
\begin{eqnarray}\label{Jkcore}
J_{{\bf k}}\left( \omega \right)&=&\left| \left\langle \widetilde{{\bf k}}%
\left| \Delta \right| c\right\rangle \right| ^{2}\int_{-\infty }^{\infty }%
\frac{dt}{2\pi }e^{i\left( \omega -\varepsilon _{{\bf k}}+E_{c}\right)
t} \\ \nonumber
&\times&\exp \left( \frac{1}{\pi }\int g_{QM}\left( \omega ^{\prime }\right)
\left( e^{-i\omega ^{\prime }t}-1\right) d\omega ^{\prime }\right) .
\end{eqnarray}
We have used Eq. \ref{tau5}, and replaced the Bloch functions with plane
waves, to find 
\begin{eqnarray}\label{gQM}
&&g_{QM}\left( \omega \right)=\\ \nonumber
&&\sum_{q}\left| \frac{-V^{q}\left( z_{c}\right)
}{\omega _{q}}+\frac{1}{i\kappa }\int_{-\infty }^{z_{c}}dze^{i\left(
\widetilde{k}-\kappa \right) \left( z-z_{c}\right) }V^{q}\left( z\right)
\right| ^{2}
\delta \left( \omega -\omega _{q}\right) .
\end{eqnarray}
Here $z_{c}$ is the position of the atom from which the photoelectron comes, 
$q$ labels the fluctuation potentials $V^{q}$ (mostly taken from Eq. \ref{Vq}%
), $q=\left( q_{z},{\bf Q}\right) $, and 
\[
\ \widetilde{k}^{2}=k_{z}^{2}+2\left( \left| V_{0}\right| +i\Gamma
_{f}\right) ,
\]
\[
\kappa ^{2}=k_{z}^{2}+2\left( \left| V_{0}\right| +\omega
_{q}+i\Gamma _{2}\right) -\left| {\bf Q}\right| ^{2}. 
\]
Further $V_{0}=\left\langle V_{eff}\right\rangle _{average}$, $V_{eff}$ is
the effective crystal potential, $\Gamma _{f}=\left| 
\mathop{\rm Im}%
\Sigma \left( \varepsilon _{{\bf k}}+\left| V_{0}\right| \right) \right| $,
and $\Gamma _{2}=\left| 
\mathop{\rm Im}%
\Sigma \left( \varepsilon _{{\bf k}}+\left| V_{0}\right| +\omega _{q}\right)
\right| $ (c. f. Eq. \ref{heff}). The photoelectron is taken to leave the
solid at right angle to the surface.

\begin{figure}
\vspace*{6.8cm}
\includegraphics{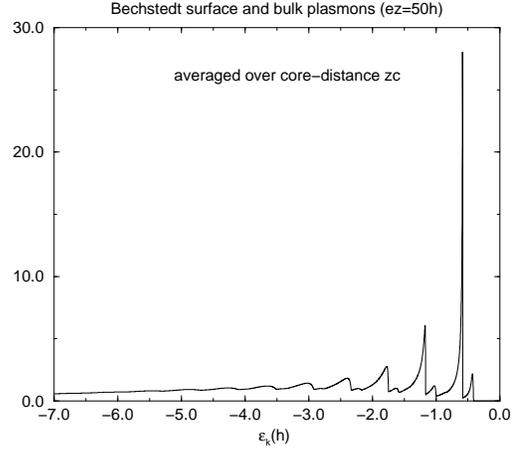}
\caption{Satellite spectra from the exponential
expression in Eq. \ref{Jkcore} in a plasmon pole approximation neglecting
electron-hole contributions but including both surface and bulk plasmons.
The spectra are averaged over core hole position and for a high photon
energy ($ez=50$).
}
\end{figure}

In the high energy limit the interference terms in Eq. \ref{gQM} drop out
and we have \cite{Hedin98}, \cite{AH83} 
\begin{equation}
g_{BS}\left( \omega \right) =\frac{\alpha \left( \omega \right) }{\omega }%
+z_{c}\tau \left( \varepsilon ,\omega \right) ,  \label{gBS}
\end{equation}
where $\alpha \left( \omega \right) $ is the function giving the intrinsic
spectrum ($\alpha \left( 0\right) $ is the MND singularity index \cite
{Mahan81}), and $\tau \left( \varepsilon ,\omega \right) $ is the
differential inverse mean free path (see e. g. Ref. \cite{Tung77}), 
\[
\tau \left( \varepsilon ,\omega \right) =\frac{1}{\pi \varepsilon }\int 
\frac{dq}{q}%
\mathop{\rm Im}%
\left[ \frac{-1}{\varepsilon \left( q,\omega \right) }\right] .
\]
Eq. \ref{gBS} gives a result roughly the same as that obtained from the
transport theory used by Penn. \cite{Penn77} \ Finally the semi-classical
theory with the photoelectron as a perturbing charge on a straight
trajectory gives \cite{Hedin98} 
\begin{eqnarray}\label{gSC}
&&g_{SC}\left( \omega \right)= 
\\ \nonumber
&&\sum_{q}\left| \frac{-V^{q}\left( z_{c}\right) 
}{\omega _{q}}+\frac{1}{iv}\int_{-\infty }^{z_{c}}dze^{-i\omega _{q}\left(
z-z_{c}\right) /v}V^{q}\left( z\right) \right| ^{2}\delta \left( \omega
-\omega _{q}\right) ,  
\end{eqnarray}
where $v$ is the photoelectron velocity.

\begin{figure}
\vspace*{7.cm}
\includegraphics{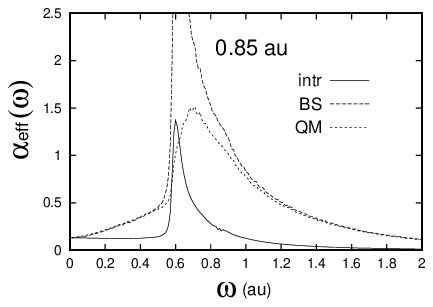}
\caption{Curves of $\alpha _{eff}\left( \omega \right) $ and $P_{k}\left(
\omega \right) $ for Al metal with a photoelectron energy of $0.85\;au$
(upper panel) and $9.85\;au$ (lower panel). The $P_{k}\left( \omega \right)$
curves are for given $k$ values as functions of $\omega $, i. e. they are
''constant final state spectra'' (CFS).
}
\end{figure}

\begin{figure}
\vspace*{6.7cm}
\includegraphics{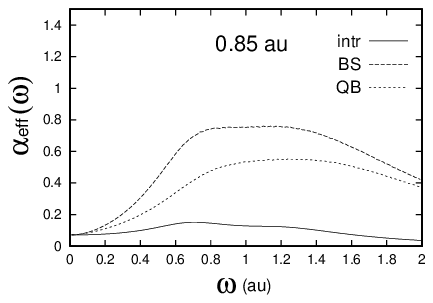}
\caption{The same results as in Fig. 7 for Cu metal.
}
\end{figure}

In Fig. 5 we show results for the bulk plasmon satellite in the three
approximations. The results are from the lowest order term in expanding the
exponential in Eq. \ref{Jkcore} using a plasmon pole approximation (no
electron-hole pairs and no surface plasmons). Actually the results from the
exponential expression and lowest order theory are identical apart from a
constant factor in the energy range up to the second satellite peak. The
semi classical theory works almost perfectly except at lower photon energies
(see curve with $ez=2$), while the Berglund and Spicer transport theory
works poorly until the energy is very high (of order keV). The quantity $ez$
is the maximum energy that the photoelectron can have with a given photon
energy.

In Fig. 6 we show the satellite region over an extended energy range at a
high photon energy ($ez=50$), calculated with the exponential expression in
Eq. \ref{Jkcore}, and averaged over core hole position. We see the strong
bulk plasmons preceded by the weaker surface plasmons. The higher order
plasmon peaks become broader and broader. The very sharp first and second
order peaks will become broader when plasmon damping and electron-hole
excitations are accounted for.

\begin{figure}
\vspace*{3.5cm}
\includegraphics{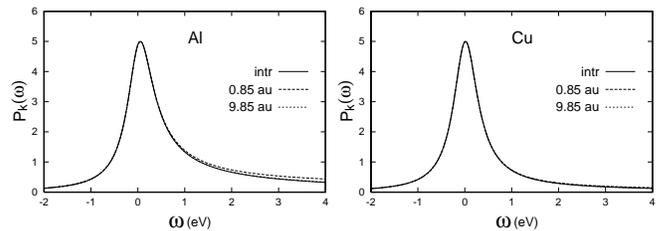}
\caption{Curves of the asymmetric quasi-particle peak for Al metal (left) and
Cu metal (right). The curves have a 300 meV Lorentzian broadening, and give
results for emission from an atom 10 au from the surface. Results for a
photoelectron kinetic energy of 0.85 and 9.85 au, and also when only the
intrinsic contribution (independent of photoelectron energy) is included.
}
\end{figure}

Al metal is rather special in having a very sharp plasmon peak carrying most
of the loss spectral strength (see e. g. Fig 4.14 in Ref. \cite{Hüfner95}).
To study the effect of the shape of the loss function we did some
comparisons between Al and Cu. We define an ''effective'' singularity
function by $\alpha _{eff}\left( \omega \right) =\omega g_{QM}\left( \omega
\right) $ which includes both intrinsic and extrinsic amplitudes, and thus
depends on the photoelectron energy. In Fig.7 we show $\alpha _{eff}\left(
\omega \right) $ for Al at two different photoelectron energies, and compare
results from taking only the intrinsic contribution (intr) with the BS and
QM cases. Clearly both extrinsic and interference effects are strong also at
the higher energy. In Fig. 7 we also show the photocurrent as function of
the photon energy obtained with the exponential expression, Eq. \ref{Jexp}
at the two photoelectron energies (CFS mode). The same results for Cu are
shown in Fig. 8. It is clear that extrinsic and interference effects are
just as strong in Cu as in Al. The difference in the loss spectra of the two
metals however shows in the shape of the satellites, the Cu satellite is
featureless as could be expected. In Fig. 9 we show both the intrinsic
(intr) and the quantum mechanical (QM) results for the quasi-particle peaks
for Al and Cu, broadened by a Lorentzian of FWHM=600 meV, and for two
photoelectron energies. It is clear that both FWHM and asymmetry in practice
are unaffected both by photoelectron energy and by extrinsic losses.

\subsection{Strongly correlated quasi two-dimensional solids}

In many high temperature superconductors (HTC) we have layered materials
with planes of CuO which are active in superconductivity. Much theoretical
work concentrates on the electronic correlations in these quasi-2D CuO
planes, and there are many interesting relations between properties found
theoretically in purely 2D systems and experiment. To interpret
photoemission data the full 3D crystal should however be taken into account.
In a recent study of Bi2212 we find strong effects on the threshold line
shape due to low energy excitations connected with the coupling between the
planes, effects which do not appear in a pure 2D treatment. \cite{HL01} \
The effects can be seen as external if we consider photoemission from one
layer and regard the layers outside as an external system, or dominantly
intrinsic if we look at the system as three-dimensional. The effect we
discuss is different from that considered by Haslinger and Joynt \cite
{Haslinger01}, who consider ohmic losses of the photoelectron occurring
after the electron has left the solid.

\begin{figure}
\vspace*{5.5cm}
\includegraphics{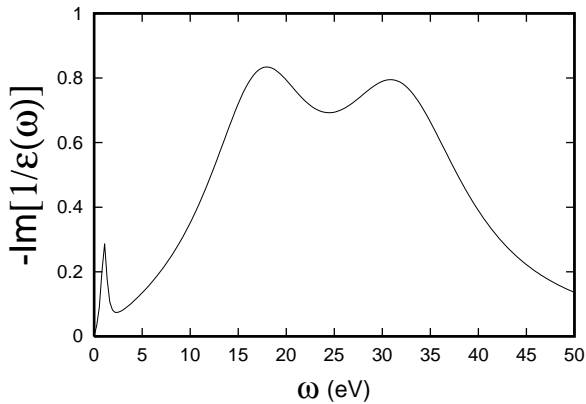}
\caption{The loss function for Bi2212 as parametrized by Norman et al.[59]
}
\end{figure}

The low energy part of the photoemission spectrum comes from exciting
electrons in a 2D layer of Bi2212. The one electron states are then quite
localized in the $z$ direction, and the shake-up and loss effects are
assumed to be similar to those in a core electron spectrum. As further
motivation for this assumption we can compare with results for simple
metals. In Sect. III we noted that the valence electron satellite structure
in simple metals is very similar to the core electron one. Further the
valence electron quasi-particle line shape, away from the Fermi level, is
markedly asymmetric just like a core line. \cite{Hedin80} \ For electrons at
the Fermi level, which is an important case, we can however not rest on a
comparison with valence electrons but have to rely on the localization in
the $z$ direction.

\begin{figure}
\vspace*{6.cm}
\includegraphics{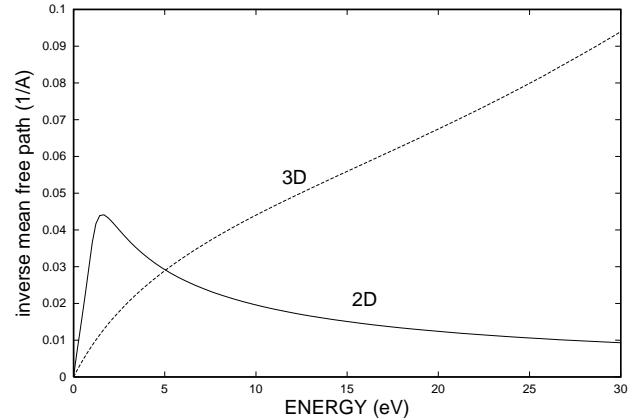}
\caption{The contributions to the inverse mean free path $%
1/\lambda $ from the 2D (full drawn) and 3D (dashed) terms in the case of
Bi2212.
}
\end{figure}

The calculations for Bi2212 take as input data a three term parametrization
(see Eq. \ref{lossfunc}) made by Norman et al \cite{Norman99} of the loss
spectrum obtained by N\"{u}cker et al \cite{Nücker89}. The parametrized
curve is shown in Fig. 10. The first sharp peak describes 2D plasma
vibrations, and the linear small $\omega $ part the acoustic plasmons coming
from coupling between the 2D layers. In Fig. 11 we show the contributions to
inverse mean free path coming from the 2D part (the first term), and the 3D
part (the second and third terms). It is clear that at about 20 eV, where
photoemission measurements usually are done, the 3D contribution dominates.
As discussed in the introduction, we take the statevectors as direct
products, and write for the initial and final states, 
\begin{equation}
\left| N_{B}\right\rangle \left| N_{2D}\right\rangle ,\;\left| N_{B}^{\ast
},s_{1}\right\rangle \left| N_{2D}-1,s_{2}\right\rangle \left| {\bf k}%
\right\rangle .  \label{eq:states1}
\end{equation}
Here $\left| N_{2D}\right\rangle $ is the state vector for the electrons in
one particular layer at a distance $z_{0}$ ($z_{0}>0$) from the surface (the
one from which the photoelectron comes), and $\left| N_{B}\right\rangle $
the state vector for the remaining (bulk) electrons which move in 3D. $%
\left| N_{2D}-1,s_{2}\right\rangle $ is an excited state $s_{2}$ of the
particular layer, and $\left| N_{B}^{\ast },s_{1}\right\rangle $ an excited
state $s_{1}$ of the bulk electrons. The star indicates that these electrons
move in the presence of a localized hole at ${\bf r}=\left( {\bf 0}%
,z_{0}\right) $, and thus is an eigenfunction of a different Hamiltonian
than that for $\left| N_{B}\right\rangle $. Finally $\left| {\bf k}%
\right\rangle $ is the photoelectron state.

Assuming that the 2D part is like a core level, we could derive an
expression for the photocurrent as a convolution between the 2D current $J_{%
{\bf k}}^{2D}\left( z_{0},\omega \right) $ and an {\it effective broadening
function} $P_{{\bf k}}\left( z_{0},\omega \right) $, 
\begin{equation}
J_{{\bf k}}\left( z_{0},\omega _{phot}\right) =\int J_{{\bf k}}^{2D}\left(
z_{0},\omega ^{\prime }\right) P_{{\bf k}}\left( z_{0},\omega _{phot}-\omega
^{\prime }\right) d\omega ^{\prime }.  \label{mainPES}
\end{equation}
A delta function peak $\delta \left( \omega -\varepsilon _{0}-\varepsilon _{%
{\bf k}}\right) $ in $J_{{\bf k}}^{2D}\left( z_{0},\omega \right) $\ will
hence give a contribution $P_{{\bf k}}\left( z_{0},\omega
_{phot}-\varepsilon _{0}-\varepsilon _{{\bf k}}\right) $ to the
photocurrent. Summing up to an exponential expression we have 
\begin{eqnarray}\label{exp3}
P_{{\bf k}}\left( \omega \right)&=&e^{-z_{0}/\lambda -a_{3D}^{intr}}\int
\frac{dt}{2\pi }
e^{-i\omega t}
\\ \nonumber
& &\times
\exp \left[ \int_{0}^{\infty }\alpha _{2D}\left( {\bf k}%
,z_{0};\omega ^{\prime }\right) \frac{e^{i\omega ^{\prime }t}-1}{\omega
^{\prime }}d\omega ^{\prime }\right.
\\ \nonumber
& & \ \ \ \ \ \  \left.+\int_{\omega _{th}}^{\infty }\alpha
_{3D}\left( {\bf k},z_{0};\omega ^{\prime }\right) \frac{e^{i\omega ^{\prime
}t}}{\omega ^{\prime }}d\omega ^{\prime }\right]. 
\end{eqnarray}
The 3D contribution is somewhat arbitrarily cut at a threshold value of $%
\omega _{th}=0.1$. We have replaced $e^{i\omega ^{\prime }t}-1$ by $%
e^{i\omega ^{\prime }t}$ in the 3D term. The reason is that in the high
energy limit we have 
\begin{equation}
\int_{\omega _{th}}^{\infty }\frac{\alpha _{3D}\left( {\bf k},z_{0};\omega
\right) }{\omega }d\omega =a_{3D}^{intr}+z_{0}/\lambda ,  \label{sumrule2}
\end{equation}
provided we neglect the contribution to the mean free path $\lambda $ coming
from 2D excitations. This, as we just remarked, is reasonable at
photoelectron energies of 20 eV and higher. Since the 2D and 3D
contributions add in an exponent we can write $P_{{\bf k}}\left(
z_{0},\omega \right) $ as a convolution, 
\begin{eqnarray*}
P_{{\bf k}}\left( z_{0},\omega \right)&=&e^{-z_{0}/\lambda
-a_{3D}^{intr}}\\ \nonumber
&\times&\int P_{{\bf k}}^{2D}\left( z_{0},\omega -\omega ^{\prime
}\right) P_{{\bf k}}^{3D}\left( z_{0},\omega ^{\prime }\right) d\omega
^{\prime }.
\end{eqnarray*}
For $P_{{\bf k}}^{3D}$ we make a Taylor expansion, and keep only the low
order result, $P_{{\bf k}}^{3D}\left( z_{0},\omega \right) =\delta \left(
\omega \right) +\alpha _{3D}\left( z_{0},\omega \right) /\omega $. We have
then omitted the multiple quasi-boson excitations starting at $\omega
=2\omega _{th}$. Since $P_{{\bf k}}^{2D}$ is normalized to unity, and
consists of a peak that is sharp compared to $\alpha _{3D}$, we can write 
\begin{equation}
P_{{\bf k}}\left( z_{0},\omega \right) \simeq e^{-z_{0}/\lambda
-a_{3D}^{intr}}\left[ P_{{\bf k}}^{2D}\left( z_{0},\omega \right) +\frac{%
\alpha _{3D}\left( {\bf k},z_{0};\omega \right) }{\omega }\right] .
\label{exp5}
\end{equation}
The extrinsic contribution to $\alpha _{2D}\left( {\bf k},z_{0};\omega
\right) $ can be neglected, and the intrinsic contribution has a $z_{0}$ but
no ${\bf k}$ dependence. There is a competition between a declining function 
$\exp (-z_{0}/\lambda )$ and an increasing function $\alpha _{2D}\left(
z_{0},\omega \right) $ making the contribution from the second layer
dominate. We approximate $\alpha _{2D}\left( z_{0},\omega \right) $ by a
rectangular function, $\alpha _{0}\theta \left( \omega _{0}-\omega \right) $
independent of $z_{0}$, and broaden with a Lorentzian of width $\Gamma $ $%
\left( FWHM=2\Gamma \right) $. As parameters we use $\alpha _{0}=0.255$ and $%
\omega _{0}=0.08$. For $\omega <\omega _{0}$ we then have the Doniach-Sunjic
expression \cite{Doniach70}, 
\begin{eqnarray}\label{Pth}
P_{{\bf k}}^{2D}\left( \omega \right)&=&C\left( \alpha _{0}\right) \frac{\cos %
\left[ \pi \alpha _{0}/2-\left( 1-\alpha _{0}\right) \arctan \left( \omega
/\Gamma \right) \right] }{\left( 1+\left( \omega /\Gamma \right) ^{2}\right)
^{\left( 1-\alpha _{0}\right) /2}},
\nonumber \\
C\left( \alpha _{0}\right) &=&\frac{%
e^{-\gamma \alpha _{0}}}{\left( \alpha _{0}-1\right) !\omega _{0}^{\alpha
_{0}}\Gamma ^{1-\alpha _{0}}\sin \left[ \pi \alpha _{0}\right] }
\end{eqnarray}
where $\gamma =0.577$ is the Euler constant. For $\omega >\omega _{0}$ $P_{%
{\bf k}}^{2D}\left( \omega \right) $ only has a weak tail with less than
10\% of the norm (for $\alpha _{0}<0.4)$. The coefficient $C\left( \alpha
_{0}\right) $ was derived in Ref. \cite{AH83}.

\begin{figure}
\vspace*{11.1cm}
\includegraphics{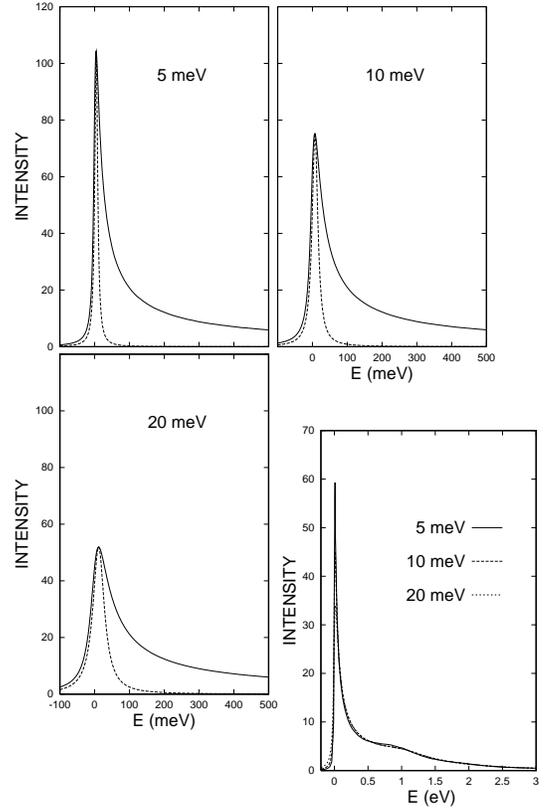}
\caption{The effective loss function $P_{k}^{2D}\left( \omega
\right) $ convoluted with Lorentzians of different widths $\Gamma $ (5, 10,
and 20 meV). Also the Lorentzians are shown. The photoelectron energy is
0.85 au.
}
\end{figure}

$P_{{\bf k}}^{2D}\left( \omega \right) $ is the function that broadens a $%
\delta $-function peak in $J_{{\bf k}}^{2D}\left( \omega \right) $. If $J_{%
{\bf k}}^{2D}\left( \omega \right) $ has a Doniach-Sunjic singular shape the
broadening with $P_{{\bf k}}^{2D}\left( \omega \right) $ still gives Eq.\ 
\ref{Pth} but with an $\alpha _{0}$ that is the sum of the alphas in $J_{%
{\bf k}}^{2D}$ and in $P_{{\bf k}}^{2D}\left( \omega \right) $. This is so
because the time transform of a power law singularity $\omega ^{-\left(
1-\alpha _{0}\right) }$ is $t^{-\alpha _{0}}$, and a convolution in
frequency space is a product in time space.

In Fig. 12 we show the sum for the first four layers of the 2D contributions 
$\exp \left( -z_{0}/\lambda -a_{3D}^{intr}\right) P_{{\bf k}}^{2D}\left(
\omega \right) $ broadened with different Lorentzians. The 3D terms are not
included except for the (all important) mean free path factor. In the first
three panels with a limited energy region (up to 500 meV) we have used the
rectangular approximation for the $\alpha _{2D}$ contributions. In the last
panel with a larger energy range the full evaluation was done since it is
superior to the rectangular model for energies above 0.5 eV. Also the
Lorentzians are shown to ease the estimate of the size of the asymmetries.
It is clear that we have a sizeable line asymmetry, and also a long tail
extending over several eV.

In the superconducting state the loss function should have a gap. We mimic
this gap by using a rectangular alpha function 
\begin{equation}
\alpha _{2D}\left( \omega \right) =\alpha _{0}\theta (\omega -\omega
_{sc})\theta \left( \omega _{0}-\omega \right) ,  \label{alpha2D1}
\end{equation}
still using $\alpha _{0}=0.255$ and $\omega _{0}=0.08\;au=2.2\;eV$. For the
gap $\omega _{sc}$ we take $\omega _{sc}=70\;meV$. In Fig. 13 we show the
corresponding $P_{{\bf k}}^{2D}\left( \omega \right) $ broaden with a
Lorentzian of width $\Gamma =15\;meV$. Our choice of parameters is only made
to illustrate the qualitative behavior to be expected. The curve clearly
shows the peak-dip-hump lineshape found experimentally (for a recent
reference see e g \cite{Campuzano99})

\begin{figure}
\vspace*{5.5cm}
\includegraphics{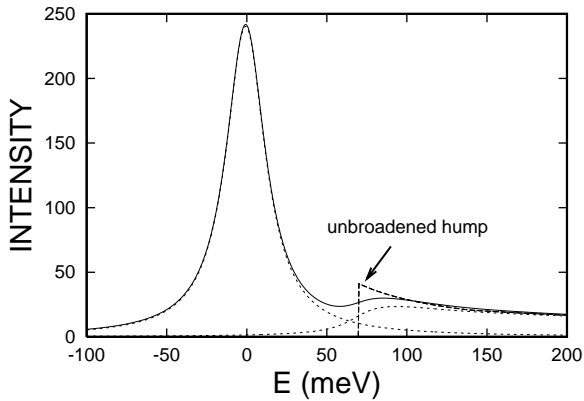}
\caption{The effective loss function $P_{k}^{2D}\left( \omega
\right) $ for a gapped spectrum using the parametrization in
Eq. \ref{alpha2D1}. The Lorentzian broadening is $\Gamma =15\;meV$.
}
\end{figure}

Recently it has been possible to obtain very accurate tunneling data from
Bi2212, and it is of interest to compare these data with the PES satellites
(Ref. \cite{note2}), since the tunneling data also show peak-dip-hump
structures. \cite{Yurgens99} PES\ and tunneling are basically different
spectroscopies. There can however be qualitative similarities since in both
cases the electrons couple to 3D quasi-boson excitations like phonons,
electron-hole pairs, plasmons, magnons etc. In our treatment of PES we take
the properties of a particular 2D layer as given and study the effect to low
order of the sudden appearance of a hole in the 2D system on the
quasi-bosons (intrinsic excitations) as well as of the coupling of the
photoelectron leaving this layer to the quasi-bosons (extrinsic
excitations), and their interference. We found that the intrinsic
contributions dominate for small excitation energies.

Tunneling is traditionally described by a spectral function which involves
matrix elements of the electron annihilation operator between the initial
state and the excited states. \cite{Schrieffer63}, \cite{Mahan81} The
excited states consist of a 2D layer state with a hole, and some state of
the quasi-bosons in the presence of a localized hole. In lowest order
perturbation theory the probability for a final state with excited
quasi-bosons is given by the first term in Eq. \ref{tau5}. This means that 
{\it the intrinsic contribution to PES and the tunneling currents are the
same }except for the mean free path effect in PES shown in Eq. \ref{exp5},
and the summation over momenta in tunneling giving the density of states,
DOS. In PES\ on the other hand, we have momentum conservation from the
dipole matrix element. As mentioned above, we modify this analysis valid for
the normal state, by simply assuming that the loss function should have a
gap in the superconducting state.

In $Bi2212$ we have a van Hove singularity (VHS) at the Fermi level, which
makes the difference between DOS and momentum conservation of less
importance (there might actually even be two VHS if the two $CuO$ planes at $%
3\;A$ apart produce a significant splitting). More important is that in PES
the electrons come from a thin surface region (of the order of the mean free
path) while in tunneling they may come from an extended region which can be
hundreds of $A$, and that the coupling functions $V(z)$ have a slow approach
to their bulk limit (Fig. 3). Additionally there are two energy gaps
(superconducting gap and pseudogap), which further complicates the picture.
It is clear that PES and tunneling structures cannot be quantitatively the
same, but since the same quasi-bosons are involved, there may well be
qualitative similarities even though the coupling strengths can be
different. We remind that we have taken the spectral function for the 2D
system (the function often calculated by theoreticians using say a t-J
model) as a sharp peak, and any peak structure in $J_{2D}\left( \omega
\right) $ has to be convoluted with the effective broadening function $%
P_{2D} $. In our analysis we have only treated the plasmons for the simple
reason that the experimental loss data at hand did not have resolution
enough to show phonons and other low energy excitations.

\section{Concluding remarks}

When we go beyond the common procedure of comparing the spectral function
with photoemission, there are a number of definitions and distinctions that
have to be introduced. In the {\it one-step approximation} only the
quasi-particle peak and the dipole matrix element are included but no
intrinsic or extrinsic losses. In the {\it intrinsic approximation} the full
spectral function including its tail from intrinsic losses, as well as the
dipole matrix element are included, but no extrinsic losses. In the {\it %
sudden approximation} (or {\it Berglund-Spicer} model) the spectral function
is taken as a source term in a transport problem, or roughly equivalently we
make a convolution between spectral function and the extrinsic loss
function. The dipole matrix element is taken as independent of loss energy.
The {\it adiabatic approximation }somehow describes the behavior at
threshold, but there is no simple approximation scheme to do this. When the
losses come from extended excitations the satellite in photoemission has
contributions both from intrinsic and extrinsic {\it amplitudes}.\ A {\it %
semi-classical approximation} puts the photoelectron on a trajectory, and
the losses are calculated with the photoelectron as an external
perturbation. It should be noted that there is considerable ambiguity in the
literature concerning these definitions.

In this paper we discuss photoemission in general, we explore the limits of
the sudden approximation, and we analyze the role of extrinsic losses. It is
not a comprehensive treatment, that would require much more space, and we
concentrate on some problems of broad interest. Thus in core electron
photoemission we e.g. do not take account of the finite core electron
lifetime. This lifetime can cause spectacular effects if phonon energies are
of the same order as the lifetime width. \cite{Almbladh77} It can also cause
large effects if there e.g. are Koster-Kronig decay channels, and two nearby
discrete levels (say $2p_{1/2}\,\ $and $2p_{3/2}$).\cite{Zaanen86} There are
many resonace effects in d and f electron systems, some involving atoms on
different sites like multiatom resonance PES. \cite{Kay01} We have omitted
this large topic. For a general review see Ref. \cite{Kotani99} 

We study in some detail how the three-dimensionality of a quasi-2D solid,
Bi2212, influences its photoemission. We have earlier found that for a
strongly correlated {\it localized} system the sudden approximation is
reached rather quickly, at about $10\;eV$ \cite{Lee99}. For a weakly
correlated system on the other hand, like an sp-metal or semiconductor, the
sudden limit in core electron photoemission is approached very slowly, on
the keV scale. \cite{Hedin98} \ The slow approach is connected with strong
destructive interference between the intrinsic and extrinsic channels for
plasmon production. The cancellation is particularly strong for small
momentum plasmons where the long-wave plasmons are excited by the average
potential from the core hole and photoelectron, which is zero \cite{Gadzuk77}%
. When we compare the satellites of Al and Cu we find that the latter have
about the same strength but are structureless. The asymmetric lineshape in
core electron photoemission from metals is, on the other hand, hardly
affected by the external loss processes \cite{Hedin98}.

We are interested in energies where the sudden limit is reached for the
strongly correlated layer from which the photoelectron comes, and derive an
expression for the photocurrent as a convolution of the sudden approximation
for the current from the layer with an effective loss function, $P_{{\bf k}%
}\left( \omega \right) $ (Eq. \ref{mainPES}). We assume, as far as the loss
properties are concerned, that the photoelectron comes from a localized
position. In our specific example, $Bi2212$, the $c$ value is $15.4\;A$
(neglecting crystallographic shear), and almost all contributions come from
the first unit cell. The two first $CuO$ layers are at $0.39\;c$ and $%
0.61\;c $ from the surface (which is between two $BiO$ layers). \cite
{Takahashi89}

To obtain $P_{{\bf k}}\left( \omega \right) $ we use a previously developed
method based on a quasi-boson\ model, where the electron-boson coupling is
given by fluctuation potentials related to the dielectric response function 
\cite{Hedin99}. We find that the energy loss function, which we take from
experimental data, can be related to the screened potential which we need to
calculate the (intrinsic and extrinsic) losses in photoemission. The
fluctuation potentials related to the electrons in the layers are universal
functions, which are easily calculated. They have some resemblance to a
surface plasmon potential, but penetrate the whole solid and have the Bloch
wave symmetry. We use the real part (or equivalently the imaginary part) of
a phase shifted bulk potential to get a potential which is zero at the
surface, and mimics the potential we have in a finite solid.

We have extrapolated results from electron energy loss experiments for q=0
to finite vales of q. This inclusion of dispersion makes the mean free path
considerably longer than obtained by Norman et al \cite{Norman99}, about $%
12\;A$ rather than $3\;A$, at say $20\;eV$ (Fig. 2). Measurements by the
ITR-2PP technique \cite{Nessler98} give a lifetime of $\tau =10\;fs$ at an
energy $\varepsilon =3\;eV\,$\ above the Fermi surface. The mean free path
is $\lambda =v\tau $. Converting energy to velocity by $mv^{2}/2=\varepsilon 
$ gives a mean free path $\tau =103\;A$ as compared to our result of about $%
17\;A$ at that energy. This is an indication that our values rather are on
the low side. It is however hard to know what is the correct conversion
between energy and velocity at such low energies, which makes a comparison
very uncertain.

From Fig. 11 we see that the 2D losses occur only for small energies, at $%
5\;eV$ the bulk losses take over. The 2D losses go to zero quite slowly,
just like the bulk losses, but on another energy scale. If we only had 2D
losses, the minimum mean free path would be long, about $20\;A$. The general
behavior of the 3D\ mean free path follows a well known pattern. The mean
free path has a minimum of about $5\;A$ at an energy of 3-4 times the energy
where the loss function has its center of gravity. We have used the Born
approximation to evaluate the mean free paths. This may seem a very crude
approximation at low energies. However the Born scattering expression with a
basis of Bloch waves and Bloch energies rather than plane waves and free
electron energies agrees with the GW approximation, which is commonly used
also at low energies. Further it was shown by Campillo et al \cite
{Campillo99} that plane waves and free electron energies was not that bad,
as long as the energies in the dielectric function are well approximated.

Our main concern is the behavior of the effective broadening function at
small energies where it is dominated by the 2D losses. The 3D contributions
set in at somewhat higher energies, and give a rather structure-less
contribution. What we here for convenience call 2D losses is of course
actually also a 3D effect since it comes from excitations of a coupled set
of 2D layers. To allow a qualitative discussion we represent the $\alpha
_{2D}$ functions by a rectangular distribution. The rectangular distribution
allows an analytic solution valid out to the cut-off $\omega _{0}$ ($\omega
_{0}\simeq 0.08\;au\simeq 2.2\;eV$). $P_{{\bf k}}\left( \omega \right) $ has
only a fairly small tail beyond $\omega _{0}$. In Fig. 12 we plot the total $%
P_{{\bf k}}\left( \omega \right) $ function (sum over the four first layers,
properly mean free path weighted), calculated with the rectangular
approximation and broadened with Lorentzians of different widths. We note
the marked asymmetry. If the $J_{2D}$ function has a power law singularity
with singularity index $\alpha _{L}$, the photoemission will have a
singularity index $\left( \alpha _{0}+\alpha _{L}\right) $.

In a paper by Liu, Anderson and Allen from 1991 \cite{Liu91}, they discussed
the lineshapes of $Bi_{2}Sr_{2}BaCu_{2}O_{8}$ along the $\Gamma -X$
direction obtained by Olsen et al \cite{Olson90} for $22\;eV$ photons. They
concluded that neither the Fermi liquid nor the marginal Fermi liquid
theories could fit the slow fall-off of the spectrum at higher energies. Our
results offer a possibility that the slow falloff may be due to intrinsic
creation of acoustic plasmons in a coupled set of $CuO$ layers, an effect
not present if only one $CuO$ layer is considered. This broadening is mostly
intrinsic, i e if we treat a 3D system we have an almost intrinsic effect.
However most theoretical discussions concern an isolated 2D system, compared
to which we find an appreciable extra broadening from the coupling between
the layers.

The PES spectra change strongly when we go to the superconducting state. The
main peak sharpens and a peak-dip-hump structure develops. This effect has
been interpreted as a coupling of the 2D\ state to the $\left( \pi ,\pi
\right) $ collective mode. \cite{Campuzano99} \ Here we find that this
effect also can arise from the gapping of the loss function caused by the
lack of low energy excitations in a superconductor as shown in Fig. 13.
Without a more accurate model we find it difficult to decide which is the
correct explanation, possibly it could be a combination of both mechanisms.
Since the gapping of the loss function is related to the superconducting
gap, also with our mechanism the hump will scale with the gap. It is clear
that the experimental peak-dip-hump structure rides on a background which is
not predicted by our expressions, nor by anyone else's. Our theory is
however rather schematic with its strict separation of a 2D and a 3D part,
while in reality the bands are hybridized. If we extend our approach to a
more detailed treatment of the underlying bandstructure, the background
could well be strongly changed. Such an extension represents a very large
numerical task, but with the present pilot treatment we can at least start
thinking seriously about the difficult background problems in photoemission.

In a recent paper Haslinger and Joynt \cite{Haslinger01} discussed a
broadening mechanism due to the interaction between the photoelectron when
outside the solid and the electrons in the solid. This is a different
mechanism than in this paper, which adds additional broadening. This
mechanism has also been discussed by Schulte et. al.\cite{Schulte01}

It should be stressed that we cannot claim any high quantitative accuracy.
We have put in dispersion in the loss function using a crude approximation.
Since however dispersion is very important we think our predictions are
substantially better than if dispersion had been neglected. We have only
considered normal emission where the electrons come from the $\Gamma $
point, while the interesting experiments concern electrons from the Fermi
surface. However there is no reason that the effective loss function should
change qualitatively when we go away from normal emission. The behavior of
the loss function when $\omega \rightarrow 0$ has been disputed. Most
authors seem to believe the approach is linear, but there are also claims
that it should be quadratic. \cite{Bozovic90} If it were quadratic, the
corresponding $\alpha $-function would start linearly rather than with a
constant. However $\alpha \left( \omega \right) $ would have to rise very
fast to reproduce the behavior of the loss function for the (quite small)
energies where it is known to be approximately linear. Thus the pure power
law behavior of $P_{k}\left( \omega \right) $ would be lost, but Lorentzian
broadened curves would probably not differ much. Our fluctuation potentials
are obtained by phase-shifting bulk potentials to make them zero at the
surface, and define them as zero outside the solid. This procedure turned
out to be fairly good in the metallic case, where we could check with more
accurately calculated fluctuation potentials. Again this approximation is
crude, but we believe it to be fundamentally better than if we had used a
step function on the bulk potential. Since the phase of the bulk potential
is arbitrary, such a procedure would anyhow have been arbitrary. To
calculate more accurate potentials\ is a very large numerical undertaking.

One may also question the use of a bulk expression to estimate of the mean
free path at the fairly low energies that we are concerned with, after all
we found strong effects when \ modifying the fluctuation potentials for
surface effects. It is not easy to make a strong statement here, and we can
only refer to ''the state of the art'', that bulk mean free paths are
successfully used in LEED and also in low energy life time calculations
which are compared with time-resolved two-photon PES (TR-2PPE) experiments. 
\cite{Echenique00}

\bigskip

\noindent
Acknowledgement - 
We thank J.W. Allen, J.C. Campuzano, A. Fujimori, A.J. Millis, M.R. Norman,
and Z.-X. Shen for constructive and informative comments. One of the authors
(J. D. L.) acknowledges the fellowship from the Japan Society for the
Promotion of Science.

\section{Appendix A}

The quasi-boson parameters are related to the dynamically screened potential 
$W({\bf r},{\bf r}^{\prime };\omega )$. Using its spectral resolution we can
write $W$ in terms of the fluctuation potentials $V^{q}$, 
\begin{eqnarray}\label{eq:Wscr1}
W\left( {\bf r},{\bf r}^{\prime };\omega \right)&=&\int \frac{\varepsilon
^{-1}({\bf r},{\bf r}^{\prime \prime };\omega )}{\left| {\bf r}^{\prime
\prime }-{\bf r}^{\prime }\right| }d{\bf r}^{\prime \prime }
\\ \nonumber
&=&\frac{1}{\left|
{\bf r}-{\bf r}^{\prime }\right| }+\sum_{q}\frac{2\omega _{q}V^{q}\left(
{\bf r}\right) V^{q}\left( {\bf r}^{\prime }\right) }{\omega ^{2}-\omega
_{q}^{2}},  
\end{eqnarray}
where,
\begin{equation}
V^{q}\left( {\bf r}\right) =\int \frac{\rho ^{q}({\bf r}^{\prime })d{\bf r}%
^{\prime }}{\left| {\bf r}-{\bf r}^{\prime }\right| },\;\rho ^{q}({\bf r}%
)=\left\langle N,q\left| \rho _{op}\left( {\bf r}\right) \right|
N\right\rangle ,  \label{eq:Vfluctpot}
\end{equation}
\[
\rho _{op}\left( {\bf r}\right) =\int \psi ^{\dagger }(x)\psi \left(
x\right) d\xi ,\;\omega _{q}=E(N,q)-E(N,0)\;-i\delta ,
\]
and $x$ stands for three space coordinates ${\bf r}$, and $\xi $ for a spin
coordinate. Since $\omega _{q}$ and $V^{q}$ are not known exactly we use the
RPA where 
\begin{equation}
V^{q}({\bf r})=\int W({\bf r,r}^{\prime };\omega _{q})\widetilde{\phi }%
_{q}\left( {\bf r}^{\prime }\right) d{\bf r}^{\prime },
\label{eq:Vfluctpot2}
\end{equation}
with $\omega _{q}=\varepsilon _{k}-\varepsilon _{l}$ ($\varepsilon _{k}>\mu
>\varepsilon _{l})$ and $\widetilde{\phi }_{q}\left( {\bf r}\right) =\phi
_{k}\left( {\bf r}\right) \phi _{l}\left( {\bf r}\right) ,$ i e the state $q$
corresponds to a particle-hole excitation. Undamped plasmons represent a
singular case, which requires a special treatment. \cite{Hedin98}, \cite
{Hedin99} \ In our treatment we will take the state $"q"$ as a quasi-boson
state. In our applications to photoemission the fluctuation potentials $V^{q}
$ always appear in a quadratic form together with $\omega _{q}$ such that
they can be eliminated and replaced with the imaginary part of the screened
potential $%
\mathop{\rm Im}%
W$%
\begin{equation}
\mathop{\rm Im}%
W\left( {\bf r},{\bf r}^{\prime };\omega \right) =-\pi \sum_{q}V^{q}\left( 
{\bf r}\right) V^{q}\left( {\bf r}^{\prime }\right) \delta \left( \omega
-\omega _{q}\right) ,\;\omega >0.  \label{ImW}
\end{equation}
$%
\mathop{\rm Im}%
W$ in turn can be estimated from experimental electron energy loss data.

\section{Appendix B}

To find the ground state for the Hamiltonian in Eq. \ref{eq:elbosham} we
replace $c_{l}^{\dagger }c_{l^{\prime }}^{{}}$ by $\delta _{ll^{\prime
}}-c_{l^{\prime }}^{{}}c_{l}^{\dagger }$ to have 
\begin{eqnarray}\label{eq:Hsyst1}
H_{syst}&=&\sum_{l}\varepsilon _{l}c_{l}^{\dagger }c_{l}^{{}}+\sum_{q}\omega
_{q}a_{q}^{\dagger }a_{q}^{{}}\\ \nonumber
&-&\sum_{qll^{\prime }}V_{ll^{\prime
}}^{q}c_{l^{\prime }}^{{}}c_{l}^{\dagger }\left[ a_{q}^{{}}+a_{q}^{\dagger }%
\right] +\sum_{ql}V_{ll}^{q}\left[ a_{q}^{{}}+a_{q}^{\dagger }\right] .
\end{eqnarray}
The last term can be combined with the quadratic boson term, and the linear
boson operator eliminated by a shift, 
\[
a_{q}=\widetilde{a}_{q}^{{}}-\frac{V_{0}^{q}}{\omega _{q}}%
,\;V_{0}^{q}=\sum_{l}V_{ll}^{q},
\]
to obtain 
\begin{eqnarray}\label{eq:Hsyst2}
H_{syst}&=&\sum_{l}\varepsilon _{l}c_{l}^{\dagger }c_{l}^{{}}+\sum_{q}\omega
_{q}\widetilde{a}_{q}^{\dagger }\widetilde{a}_{q}^{{}}\\ \nonumber
&-&\sum_{qll^{\prime
}}V_{ll^{\prime }}^{q}c_{l^{\prime }}^{{}}c_{l}^{\dagger }\left[ \widetilde{a%
}_{q}^{{}}+\widetilde{a}_{q}^{\dagger }-2\frac{V_{0}^{q}}{\omega _{q}}\right]
-\sum_{q}\frac{\left( V_{0}^{q}\right) ^{2}}{\omega _{q}} 
\end{eqnarray}
The last term in the square bracket is combined with the first term, which
then is diagonalized. This gives 
\begin{eqnarray*}
H_{syst}&=&\sum_{l}\widetilde{\varepsilon }_{l}\widetilde{c}_{l}^{\dagger }%
\widetilde{c}_{l}^{{}}+\sum_{q}\omega _{q}\widetilde{a}_{q}^{\dagger }%
\widetilde{a}_{q}^{{}}\\ \nonumber
&-&\sum_{qll^{\prime }}\widetilde{V}_{ll^{\prime }}^{q}%
\widetilde{c}_{l^{\prime }}^{{}}\widetilde{c}_{l}^{\dagger }\left[
\widetilde{a}_{q}^{{}}+\widetilde{a}_{q}^{\dagger }\right] +\sum_{q}\frac{%
\left( V_{0}^{q}\right) ^{2}}{\omega _{q}}.
\end{eqnarray*}
The ground state for $H_{syst}$ is thus a filled Fermi sea with no
quasi-bosons $\widetilde{a}_{q}^{{}}$, i. e. a simple Slaterdeterminant $%
\left| N^{IP}\right\rangle $, and its energy is $E_{0}=\sum_{l}\widetilde{%
\varepsilon }_{l}+\sum_{q}\left( V_{0}^{q}\right) ^{2}/\omega _{q}$. This is
also the ground state for $H$ since it does not contain any photoelectron
and thus both $h$ and $V$ give zero. The transformation matrix that
diagonalizes the $c_{l}^{\dagger }c_{l}^{{}}$ term deviates from $\delta
_{ll^{\prime }}$ by terms of order $\left( V^{q}\right) ^{2}$, and we will
neglect the difference between $\widetilde{V}_{ll^{\prime }}^{q}$ and $%
V_{ll^{\prime }}^{q}$ and between $\widetilde{\varepsilon }_{l}$\ and $%
\varepsilon _{l}$. When the occupied state $l$ is a core state the
transformation matrix is unity and $\widetilde{V}_{ll^{\prime
}}^{q}=V_{ll^{\prime }}^{q}$.

The state $c_{l}\left| N^{IP}\right\rangle $ appearing in Eq. \ref{tau3} is
not an eigenstate of $H_{syst}$. We note that states $\left\{ c_{l}\left|
N^{IP}\right\rangle \right\} ,\;\left\{ \widetilde{a}_{q}^{\dagger
}c_{l}^{{}}\left| N^{IP}\right\rangle \right\} ,\;\left\{ \widetilde{a}%
_{q^{{}}}^{\dagger }\widetilde{a}_{q^{\prime }}^{\dagger
}c_{l^{{}}}^{{}}\left| N^{IP}\right\rangle \right\} $ form a complete set to
span $\left| N-1,s\right\rangle .$ We have to lowest order in the coupling
functions $V^{q}$ 
\[
\left| l\right\rangle =\left| l\right\rangle ^{0}+\sum_{ql^{\prime }}\frac{%
V_{ll^{\prime }}^{q}}{\omega _{q}-\varepsilon _{l^{\prime }}+\varepsilon _{l}%
}\left| ql^{\prime }\right\rangle ^{0}
\]
where 
\[
\left| l\right\rangle ^{0}=c_{l}\left| N^{IP}\right\rangle ,\;\left|
ql\right\rangle ^{0}=\widetilde{a}_{q}^{\dagger }c_{l}^{{}}\left|
N^{IP}\right\rangle ,
\]
We express the ''initial state'' $\left| l\right\rangle ^{0}$ in eigenstates
of $H_{syst}$, 
\[
\left| l\right\rangle ^{0}=\left| l\right\rangle -\sum_{ql^{\prime }}\frac{%
V_{ll^{\prime }}^{q}}{\omega _{q}-\varepsilon _{l^{\prime }}+\varepsilon _{l}%
}\left| ql^{\prime }\right\rangle .
\]
We need the matrix elements $T_{s}$ (c.f. Eq. \ref{tau3}), 
\[
T_{s}=\left\langle N-1,s\left| \left( 1+V\frac{1}{E-H_{syst}-h-QVQ}\right)
c_{l^{\prime }}\right| N^{IP}\right\rangle .
\]
We are interested in the two cases when $"s"$ corresponds to $\left|
l\right\rangle $ and to $\left| ql\right\rangle $, 
\[
\left\{
\begin{array}{c}
\left| N-1,s\right\rangle =\left| l\right\rangle  \\
E\left( N-1,s\right) =E\left( N\right) -\varepsilon _{l}
\end{array}
\right. ,
\]
\[
\left\{
\begin{array}{c}
\left| N-1,s\right\rangle =\left| ql\right\rangle  \\
E\left( N-1,s\right) =E\left( N\right) -\varepsilon _{l}+\omega _{q}
\end{array}
\right. .
\]
We have, neglecting $O\left( V^{2}\right) $ terms, 
\[
T_{l}=\left\langle l\left| \left( 1+V\frac{1}{E-H_{syst}-h-QVQ}\right)
\right| l^{\prime }\right\rangle ^{0}=\delta _{ll^{\prime }}
\]
\begin{eqnarray*}
T_{ql}&=&\left\langle ql\left| \left( 1+V\frac{1}{E-H_{syst}-h-QVQ}\right)
\right| l^{\prime }\right\rangle ^{0} \\ \nonumber
&=&\frac{-V_{l^{\prime }l}^{q}}{\omega _{q}-\varepsilon _{l}+\varepsilon
_{l^{\prime }}}+\delta _{ll^{\prime }}V^{q}\frac{1}{\omega _{q}+\varepsilon
_{{\bf k}}-h-\Sigma \left( \omega _{q}+\varepsilon _{{\bf k}}\right) }.
\end{eqnarray*}
Combining these results with Eqs.\ref{Jk} \ and \ref{tau3}\ we have, 
\begin{eqnarray}\label{Jk3}
J_{{\bf k}}\left( \omega \right)&=&\sum_{l}\left| \tau _{l}\left( {\bf k}%
\right) \right| ^{2}\delta \left( \omega -\varepsilon _{{\bf k}}+\varepsilon
_{l}\right)\\ \nonumber
&+&\sum_{ql}\left| \tau _{ql}\left( {\bf k}\right) \right|
^{2}\delta \left( \omega -\omega _{q}-\varepsilon _{{\bf k}}+\varepsilon
_{l}\right)   
\end{eqnarray}
with 
\[
\tau _{l}\left( {\bf k}\right) =\left\langle \widetilde{{\bf k}}\left|
\Delta \right| l\right\rangle 
\]
\begin{eqnarray}\label{tau4}
\tau _{ql}\left( {\bf k}\right)&=&\sum_{l^{\prime }}\frac{-\left\langle 
\widetilde{{\bf k}}\left| \Delta \right| l^{\prime }\right\rangle
V_{l^{\prime }l}^{q}}{\omega _{q}-\varepsilon _{l}+\varepsilon _{l^{\prime }}%
}\\ \nonumber
&+&\left\langle \widetilde{{\bf k}}\left| V^{q}\frac{1}{\omega
_{q}+\varepsilon _{{\bf k}}-h-\Sigma \left( \omega _{q}+\varepsilon _{{\bf k}%
}\right) }\Delta \right| l\right\rangle .  
\end{eqnarray}
Here the first term is the intrinsic and the second the extrinsic amplitude.
For photoemission from a core level only one state is involved, and we can
replace all $l$ by one single index, say $c$. Our expression then reduces to
our earlier result in Refs. \cite{Bardy85} and \cite{Hedin98}. Since $\tau
_{ql}\left( {\bf k}\right) $ is quadratic in the fluctuation potentials $%
V^{q}$ Eq. \ref{tau4} can also be written in terms of $%
\mathop{\rm Im}%
W$.

\end{multicols}


\begin{references}
\bibitem{note1}  The states above \ the vacuum level have free electron
energies since in photoemission the normalization volume is always much
larger than the volume of the solid (which in turn tends to infinity). The
matrix elements in the photoemission expression however make the
quasiparticle energies enter.
\end{references}
\end{document}